\documentclass[a4paper,12pt]{article}

\pdfoutput=1 % if your are submitting a pdflatex (i.e. if you have
\usepackage{jcappub} 

\usepackage[bottom]{footmisc}

\usepackage{latexsym,array,theorem,mathrsfs,subfigure,bm,float}
\usepackage{amsmath,scalerel}
\usepackage{amsfonts}
\usepackage{amssymb}
\usepackage{xcolor}
\usepackage{bbm}

\newcommand{\bea}{\begin{eqnarray}}
\newcommand{\eea}{\end{eqnarray}}
\newcommand{\be}{\begin{equation}}
\newcommand{\ee}{\end{equation}}

\newcommand{\U}{{\rm U}}

\newcommand{\blue}{\textcolor{black}}

\newcommand{\xii}{\xi}

\newcommand{\y}{c}
\newcommand{\nn}{\nonumber}

\newcommand{\dg}{\sqrt{-g}}
\newcommand{\df}{\sqrt{-f}}
\newcommand{\dr}{\phi}
\newcommand{\h}{{ h }}
\newcommand{\hh}{{ \Psi }}

\newcommand{\al}{s}

\renewcommand{\theequation}{\arabic{section}.\arabic{equation}}

\begin{document}

\begin{flushright}
{YITP-18-87, IPMU18-0135}
\end{flushright}

\title{The Ogievetsky-Polubarinov massive gravity and the benign Boulware-Deser mode}

\author[a,b]{Shinji Mukohyama}
\emailAdd{\tt shinji.mukohyama@yukawa.kyoto-u.ac.jp}
\affiliation[a]{Center for Gravitational Physics, 
Yukawa Institute for Theoretical Physics, Kyoto University, 606-8502, Kyoto, Japan
}
\affiliation[b]{Kavli Institute for the Physics and Mathematics of the Universe (WPI), 
The University of Tokyo Institutes for Advanced Study, The University of Tokyo, Kashiwa, Chiba 277-8583, Japan
}
\author[c,d]{Mikhail~S.~Volkov}
\emailAdd{\tt michael.volkov@idpoisson.fr}
\affiliation[c]{
Institut Denis Poisson, UMR - CNRS 7013, \\ 
Universit\'{e} de Tours, Parc de Grandmont, 37200 Tours, France}
%\affiliation{Center for Gravitational Physics, Yukawa Institute for Theoretical Physics, Kyoto University, 606-8502, Kyoto, Japan
%}
\affiliation[d]{
Department of General Relativity and Gravitation, Institute of Physics,\\
Kazan Federal University, Kremlevskaya street 18, 420008 Kazan, Russia
}

\abstract{ 
\vspace{1 cm}

We  present our analysis of  the theory constructed in 1965 by 
Ogievetsky and Polubarinov (OP) --  the 
first ever theory of interacting  massive gravitons. 
Its mass term is adjusted in such a way that the non-linear field equations 
imply as a consequence the linear Hilbert-Lorentz condition, which restricts the spin 
of states in the theory. 
Strikingly, for special parameter values
this theory coincides  with one of the ``ghost-free" 
massive gravity models rediscovered only in 2010.  For generic parameter values, however, 
it propagates 6 degrees of freedom and shows ghost 
around flat space.   Surprizingly, we find that the de~Sitter space 
remains stable for a large region of the parameter space, provided that the Hubble 
expansion rate is large enough, hence 
the Boulware-Deser mode is benign in this case.
We study also other solutions and find that the Milne universe -- 
a sector of Minkowski space -- is stable in the UV limit. This presumably 
implies that at the non-linear level the ghost instability in flat space develops only 
for long waves, similarly to the classical Jeans instability. 

}

%\pacs{04.20.Fy, 04.50.Kd, 11.27.+d, 98.80.Cq}

\maketitle

\section{Introduction}
\setcounter{equation}{0}

In this paper we analyze the massive gravity theory constructed in 1965 by 
Ogievetsky and Polubarinov (OP) \cite{OP}\footnote{V. I. Ogievetsky and 
I. V. Polubarinov worked in Dubna  in the Soviet Union times.}.  
To our knowledge,  this had been the 
first serious work on massive gravity after Fierz and Pauli  \cite{Fierz:1939ix}, and 
the first ever systematic study of interacting massive gravitons. 
Among other things, OP obtained one of the ``ghost-free" massive gravity models 
rediscovered again only in 2010 \cite{deRham:2010kj}. However, their work  
is almost unknown in the modern massive gravity community, 
presumably because their strategy was quite different 
from what  is generally adopted at present.  
Therefore, 
in what follows we shall present our analysis of the OP theory and of 
some of its applications. 

To understand the OP's motivations, consider free massive gravitons in Minkowski space described 
by a symmetric tensor $\hh_{\mu\nu}$ subject to \cite{Fierz:1939ix}
\bea
(\partial^\sigma\partial_\sigma -m^2)\hh_{\mu\nu}&=&0\,,    \label{1a}   \\
\partial^\sigma \hh_{\sigma\mu}&=&0\,,        \label{2a}   \\
\hh^\sigma_{~\sigma}&=&0\,.                           \label{3a}    
\eea 
The tensor $\hh_{\mu\nu}$ has 10 independent components, but 
the five conditions in \eqref{2a} and \eqref{3a} 
eliminate the spin-1 and two spin-0 representations
and there remains only the massive spin-2
carrying  $5=10-5$ degrees of freedom (DoF). As emphasized already by Fierz and Pauli 
(FP) \cite{Fierz:1939ix}, the 
Lorentz condition \eqref{2a} is absolutely essential, because if it
were replaced by some other four conditions to keep the same number of DoF, 
the canonical energy of $\hh_{\mu\nu}$ would be non-positive. 

The OP's goal was to find a non-linear completion for Eqs.\eqref{1a}--\eqref{3a}. 
They adopted the field-theoretical approach initiated by Papapetrou \cite{Papapetrou:1948jw},
Gupta \cite{Gupta}, and Feynman  \cite{Feynman:1963ax}, and considered gravitons  
as interacting fields in {\it flat space}. Therefore, 
they kept
$\hh_{\mu\nu}$ (or rather $\hh^{\mu\nu}$) as the principle variables 
and were looking for non-linear 
terms to be added to \eqref{1a} 
to describe the graviton interactions. 

To illustrate the idea, let us consider the massless case -- the General Relativity. 
It is well known that the Einstein equations can be represented in the Papapetrou form 
(see, e.g. \cite{poisson}) as 
\be                     \label{Papa}
\partial_{\mu}\partial_\nu \left(
{\mathfrak g}^{\alpha\beta}{\mathfrak g}^{\mu\nu}-{\mathfrak g}^{\alpha\nu}{\mathfrak g}^{\beta\mu}\right)
=16\pi G\,(-g)\,
t^{\alpha\beta}_{\rm LL}\,,
\ee
where ${\mathfrak g}^{\mu\nu}=\sqrt{-g}\,g^{\mu\nu}$ and  
the Landau-Lifshitz pseudo-tensor 
$t^{\alpha\beta}_{\rm LL}$ does not contain 
second derivatives. It is always possible to impose the harmonic gauge condition  
$\partial_\mu {\mathfrak g}^{\mu\nu}=0$. Introducing the tensor $\hh^{\mu\nu}$ via 
\be                            \label{g-g}
\sqrt{-g}\,g^{\mu\nu}
=\eta^{\mu\nu}+\hh^{\mu\nu},
\ee
Eqs.\eqref{Papa} assume the form 
\bea        \label{5}
\partial^\sigma\partial_\sigma\,  
\hh^{\mu\nu}&=&(\mbox{terms non-linear in $\hh^{\mu\nu}$}),   \\
\partial_\sigma \hh^{\sigma\nu}&=&0. \label{5a}
\eea
These equations can be viewed as describing gravitons 
in Minkowski space. The non-linear terms on the right  in 
\eqref{5} describe graviton interactions.  
In the linear approximation one neglects the interaction terms 
and the equations describe free gravitons, 
\bea        \label{7}
\partial^\sigma\partial_\sigma\,  \hh^{\mu\nu}&=&0,   \\
\partial_\sigma \hh^{\sigma\nu}&=&0.  \label{7a}
\eea
One can then wonder if it is possible to go back from these linear equations 
to  the non-linear ones \eqref{5},\eqref{5a} and
apply  the field theory methods to   
reconstruct the interaction terms~? In other words, can one 
obtain the General Relativity as the
non-linear completion for the theory of free gravitons, without relying on methods of 
differential geometry~?  Today we know that this is  indeed possible  
\cite{Deser:1969wk}, but in 1965 
this fact was not known. 

Therefore,  the OP's aim was to apply the field theory methods to 
construct  non-linear terms to be added to the right hand side of \eqref{1a} to obtain 
a consistent self-interacting  theory. 
 Remarkably, they achieved  the goal
and, starting from the very first principles, constructed a fully interacting theory 
whose action contains the Einstein-Hilbert kinetic term and has also 
a graviton mass term. Sending the graviton mass to zero they recovered the General Relativity. 
Therefore, OP have been the first to obtain the General Relativity by applying only field-theory 
methods, without using the differential geometry\footnote{
The well-known paper \cite{Deser:1969wk} of 
Deser on a similar subject (considering only the massless case)  
appeared a few years after the OP's work. 
Deser used the bootstrap method, quite different from the OP's approach. }.

The central role in their construction is played by the 
subsidiary conditions \eqref{2a} and \eqref{3a}. 
However, OP had realized that it would have been technically too difficult
to keep both of them. Therefore   they 
imposed only  one combined Hilbert-Lorentz condition, 
\be                           \label{4a} 
\partial_\mu( \hh^{\mu\nu}+q\,\eta^{\mu\nu} \hh^\alpha_{~\alpha})=0,
\ee
with constant $q$. This is necessary, although not sufficient, for exclusion of negative energies. 
 They  called this condition ``spin limitation principle". 
It excludes the spin-1 and a spin-0, but not the 
second spin-0, hence there remain altogether 6 DoF. 
 OP required the formula \eqref{4a} to be 
exactly the same also in the presence of interactions: it should always contain 
partial and not covariant derivatives\footnote{This is indeed 
possible in a bimetric theory.}.  
 Therefore, the spin limitation condition always remains ``clean" 
  and removes precisely the spin-1 and spin-0 in the strict representation theory sense. 
 This is probably the most important moment: OP keep control over the spin content 
 of their theory.

%%%%%%%%%%%%%%%%%%%%%%%%%%%%
 
\blue{
 At this point, it is worth posing to compare the OP's strategy with the 
logic commonly adopted  at present, according to which 
the ``healthy" massive gravity theory has to have 5 DoF 
 to avoid the Boulware-Deser (BD)
 ghost \cite{Boulware:1973my}, hence it should contain 5 constraints \cite{deRham:2010kj}.
For a flat background\footnote{\blue{More generally, for Einstein space backgrounds.}}
these constraints have the structure similar to that in Eqs.\eqref{2a},\eqref{3a} and they eliminate 
precisely the spin-1 and two spin-0 states. However, for arbitrary backgrounds the fifth 
constraint has a rather complex structure and it is not obvious what spin states it 
eliminates\footnote{\blue{Already the linearized version 
of the 5-th constraint is very complex; see Appendix B in \cite{Mazuet:2018ysa}.  
  }}. 
This suggests that for generic backgrounds the theory may propagate
superpositions of states of different spins,
even though   the total number of DoF is always 5. 
Therefore, the theory controls the number of DoF but does not seem to always control 
their spin contents, which 
 might explain  why it shows pathologies
 for some backgrounds \cite{DeFelice:2012mx,Fasiello:2013woa,Chamseddine:2013lid}. 
}

The OP's strategy was quite different. They  constructed a theory with 6 DoF and did not care 
about the BD mode (the ghost problem was not known at the time). 
Instead, they preferred to have 
control over the spin contents of their theory -- it
contains only the spin-2 and a spin-0, while spin-1 states are definitely excluded.
Whether or not this  makes sense is to be understood.

 Getting back to their construction, 
 OP required the condition \eqref{4a} to be 
a differential consequence of the second order field equations.  
This requirement lead to certain identity relations for the Lagrangian, 
implying the existence of a local internal symmetry. By analyzing the structure of the 
symmetry generators, OP concluded that the symmetry must formally coincide with the
spacetime diffeomorphism symmetry, viewed in their approach 
as  the  {\it internal symmetry }
acting on  gravitons in flat space. OP were then able to construct 
the interaction terms order by order by requiring that the symmetry algebra closes. 
They ended up with a theory whose 
kinetic term coincides with the standard Einstein-Hilbert term for the 
``effective"  metric  $g_{\mu\nu}$ whose inverse $g^{\mu\nu}$ is 
related to the graviton field $\hh^{\mu\nu}$ via the relation similar to  \eqref{g-g},
\be                                  \label{g-g1}
\left(\frac{\sqrt{-g}}{\sqrt{-\eta}}\right)^{s+1} ((\hat{g}^{-1})^n)^{\mu\nu}=\eta^{\mu\nu}+\hh^{\mu\nu}. 
\ee 
Here the  parameters $s,n$ are {\it real} and the precise meaning of the 
matrix power will be specified below. 
Therefore, the spacetime metric 
$g_{\mu\nu}$ arises in their approach as  a secondary object related to 
the primary graviton field $\hh^{\mu\nu}$ in a very 
non-linear way, via \eqref{g-g1}. Notice that this transformation is invertible
and can be resolved with respect to $g_{\mu\nu}$.  
The OP action contains also a mass term constructed
from $g^{\mu\nu}$ and $\eta_{\mu\nu}$. 

Once the OP theory is obtained, it can be formulated 
entirely in terms of $g_{\mu\nu}$ and $\eta_{\mu\nu}$, and then it can be viewed 
simply as a bimetric theory. It implies certain on-shell conservation conditions which,
when expressed in terms of the variables $\hh^{\mu\nu}$ defined by Eq.\eqref{g-g1},
assume the form of the linear 
``spin limitation principle". 
All of this will be explained below.

Summarizing, there are two aspects of the OP's work. 
First, it presents  the first systematic derivation of the Einstein-Hilbert 
kinetic term starting from the free theory and applying 
only the field theory principles. This is, of course,  
a remarkable achievement for which OP should be fully credited, in our opinion. 
Secondly, their procedure gives also a particular mass term,  but the status of this 
is less clear, since it gives rise to 
6 DoF -- a  property considered today 
as unacceptable.  

At the same time, the OP mass term is a part of the very carefully designed  
derivation procedure.  
For a one-dimensional subset of the parameter space it shows 
the FP property and propagates only 5 DoF 
around flat space. For one particular point of this FP subset the 
theory propagates 5 DoF even at the non-linear level and 
coincides with one of the ghost-free dRGT\footnote{dRGT 
-- after the names of authors of 
\cite{deRham:2010kj}.}  
models. Therefore, the OP procedure gave  in 1965
the result that was rediscovered again only in 2010 !  

All of this suggests that the OP massive gravity deserves studying, even though 
it propagates in general 6 DoF. Therefore, we present in what follows our analysis 
of this theory and of its solutions. Skipping its derivation indicated above and described 
in the OP's paper \cite{OP}, we come directly to the theory itself. 
In modern terms, this is a bimetric theory\footnote{\blue{A bimetric theory is any theory 
with two metrics. It can be a massive gravity if only one of the
metrics is dynamical, or a bigravity if both metrics are dynamical. }
} containing the dynamical metric 
$g_{\mu\nu}$  and a non-dynamical reference metric $f_{\mu\nu}$, with a specially designed 
interaction potential constructed from these metrics.

In Section II, we rewrite  the theory in modern notation and explain 
how the linear  ``spin limitation condition" follows from the non-linear field equations. 
In brief, this is simply the condition for  the tensor obtained by varying 
the action with respect to $f_{\mu\nu}$. This tensor is conserved on-shell, which is true in any 
bimetric theory, but only in the OP theory the conservation condition can be made linear 
by the non-linear field redefinition \eqref{g-g1}. 

We then study in Section III the simplest 
solutions, such as the de Sitter or Minkowski, 
and explicitly show that there are 6 propagating DoF, 
unless for a one-dimensional subset of the parameter space for which there are 
only 5 DoF. In Section IV we obtain the effective 
action for fluctuations and, surprisingly, find that the de Sitter space 
is completely free of ghosts and gradient instabilities for a large region of the parameter 
space, although   the flat space always 
shows ghost away from the FP limit. 
This is, perhaps, our most interesting  finding -- the fact that the 6-th 
polarization can be totally harmless. 

We then proceed to study in Section V other
homogeneous and isotropic cosmologies in the theory. We find many different types
of such solutions, but unfortunately most of 
them are unstable. At the same time, it turns out that the Milne space -- 
a sector of Minkowski space -- is stable in the UV limit. This suggests  that the ghost 
instability of the flat space develops only for long waves, similarly 
to the classical Jeans instability.  Our conclusions are formulated in Section VI, 
while the Appendix contains the derivation of the stability  conditions for the 
homogeneous and isotropic cosmologies. 

Few words about the impact of the OP's work.  In the older days it 
was mentioned  in the massive gravity context \cite{vanDam:1970vg,Vainshtein:1972sx,Zeldovich:1986ft}. 
The important special case in which the theory  propagates only 5 DoF and coincides 
with the dRGT theory  was studied by Maheshwari in 1972 \cite{Maheshwari:1972mb} 
(see \cite{Pitts:2015aqa} for an interesting historical  account). 
Nowadays it is cited by experts in various field theory domains  (see for example 
\cite{Ivanov:1981wn,Cutler:1986dv,Tomboulis:1996cy,Boulanger:2000rq,%
Zinoviev:2006im,Morris:2018mhd}), but it is almost totally unknown  
to the modern massive gravity community. This has given  us the motivation for writing this text.  
%following the earlier work \cite{Freund:1968zz}. 

\section{The OP theory}
\setcounter{equation}{0}

The OP theory is a particular case of 
bimetric massive gravity. 
Any such theory is described by the dynamical metric 
$g_{\mu\nu}$ and a non-dynamical reference metric $f_{\mu\nu}$. There is no
general rule for choosing the latter, for example one can set it to be the Minkowski metric, 
$f_{\mu\nu}=\eta_{\mu\nu}$, 
but one can just as well leave it unspecified for the time being. 

The action of the theory is 
\be                 \label{act}
S=M_{\rm Pl}^2\int\left(\frac12\, R(g)-m^2 U +{\cal L}_{\rm matter} \right)\sqrt{-g}\, d^4x\,,
\ee
where $m$ is a mass parameter and 
$U$ is a scalar function constructed from products of $f_{\mu\nu}$ with 
the inverse of the physical metric, $g^{\mu\nu}$. 
Introducing the matrix $\hat{S}$ with components
\be                          \label{Sh}
\blue{  (\hat{S})^\mu_{~\nu}\equiv S^\mu_{~\nu}=g^{\mu\sigma}f_{\sigma\nu} }
 \ee
and using brackets to denote trace, $[\hat{S}]=S^\sigma_{~\sigma}$, the potential 
can be any function of traces of powers of $\hat{S}$,
\be
U=U\left([\hat{S}],[\hat{S}^2],[\hat{S}^3],\det(\hat{S})\right).
\ee
For example, it can be given by the series 
\be
U=a_0+a_1\,[\hat{S}]+a_2\,[\hat{S}^2]+a_3\,[\hat{S}]^2+\ldots,
\ee 
where $a_k$ are constant coefficients. 
This theory generically shows  
6 DoF in the gravity 
sector. If $a_2+a_3=0$ then 
the theory is said to fulfill the FP property and 
shows only 5 DoF around flat space. 
However, even then an extra 6-th polarization 
emerges when one deviates from flat space, 
\blue{unless the potential is further fine-tuned}. 
This extra mode carries a negative 
kinetic energy and is called BD ghost \cite{Boulware:1973my}. 

As a result, there exist infinitely many massive gravity theories corresponding 
to infinitely many possibilities to choose the potential $U$. One therefore needs
a guiding principle to select one particular theory. An example of this is provided 
by the dRGT potential selected by the requirement that the theory should always propagate 5 DoF,
for any backgrounds \cite{deRham:2010kj}. This potential is expressed in terms 
of fractional powers of $\hat{S}$, 
\be                   \label{dRGT}
{U}=\beta_0+\sum_{n=1}^3 \beta_k\,{U}_k\,,
\ee
where $\beta_0,\beta_k$ are real parameters and 
\bea                        \label{4}
{U}_1&=&[\hat{\gamma}],~~~~~
{U}_2
=\frac{1}{2!}([\hat{\gamma}]^2-[\hat{\gamma}^2]),~~~~~
{U}_3=
\frac{1}{3!}([\hat{\gamma}]^3-3[\hat{\gamma}][\hat{\gamma}^2]+2[\hat{\gamma}^3]),
\eea
the matrix $\hat{\gamma}$ being  determined by the condition  $\hat{\gamma}^2=\hat{S}$
hence 
$\hat{\gamma}=\hat{S}^{1/2}$. 

The OP potential is selected by a different requirement: the theory should imply the linear 
Hilbert-Lorentz condition as a consequence of the field equations. 
This potential contains arbitrary real powers of $\hat{S}$. 
Splitting the inverse metric  as
\be                      \label{split}
g^{\mu\nu}=f^{\mu\nu}+\chi^{\mu\nu}\,,
\ee
one has 
 \be                          \label{Shh}
 S^\mu_{~\nu}=g^{\mu\sigma}f_{\sigma\nu}=\delta^\mu_\nu+\chi^{\mu\sigma}f_{\sigma\nu}\equiv 
 \delta^\mu_\nu+\chi^\mu_{~\nu}\,,
 \ee
which can be written as\footnote{
\blue{We use the hat is used to denote matrices, e.g. $\hat{S}$, the matrix
components being denoted either without hat, $S^\mu_{~\nu}$, or as  $(\hat{S})^\mu_{~\nu}$.} 
}
 \be                            
 \hat{S}=\hat{g}^{-1}\hat{f}\equiv \hat{1}+\hat{\chi}\,.
 \ee
An arbitrary real power of $\hat{S}$ is defined via the series, 
 \be                           \label{Sigma}
 \hat{\Sigma}\equiv 
 \hat{S}^n=\hat{1}+n\hat{\chi}+\frac{n(n-1)}{2}\,\hat{\chi}^2+\ldots 
 \ee
Introducing the scalar 
\be                  \label{1}
\phi=\frac{1}{\sqrt{{\det} ({\hat S})}}=
\left(
1+[\hat{\chi}]+\frac12( [\hat{\chi}]^2-[\hat{\chi^2}])+\ldots
\right)^{-1/2}\,,
\ee
the OP mass term  is given by
  \bea             \label{pot}
 U&=&{\cal U} +\lambda_g \, ,                   
 \eea
where the $\hat{S}$-dependent part is 
 \be                           \label{U}
 {\cal U}=\frac{1}{4n^2}\,\dr^\al\, [\hat{S}^n]=\frac{1}{4n^2}\left(\det({ \hat S})\right)^{-s/2}[\hat{S}^n],
 \ee
while the constant part is
 \be
  \lambda_g=\frac{n-2\al-2}{2n^2}.
 \ee
The potential depends on two real parameters 
$n,\al$  (OP use instead $n,p=-(\al+1)/n$). 

The OP theory propagates 6 DoF for generic values of $n$ and $\al$. 
If the parameters belong to the ellipse shown in Fig.\ref{Fig1} below, 
then the theory has the FP property and propagates only 5 DoF around flat space. 
For the particular point at the ellipse corresponding to 
\be
s=0,~~~n=\frac12, 
\ee
the OP potential \eqref{pot} coincides with 
the dRGT potential \eqref{dRGT} for
\be
\beta_0=-3, ~~~\beta_1=1,~~~ \beta_2=\beta_3=0.
\ee
The OP theory becomes ``ghost-free" in this case, in the sense 
that it propagates only 5 DoF for generic backgrounds.

The matter term in the action \eqref{act} can be arbitrary, but we shall be considering 
just the (boldfaced) cosmological constant,
\be           \label{mat}
{\cal L}_{\rm matter}=-\bm{\Lambda}.
\ee
 
\subsection{Field equations}

 Let us vary the two metrics, $g_{\mu\nu}\to g_{\mu\nu}+\delta  g_{\mu\nu}$ and 
 $f_{\mu\nu}\to f_{\mu\nu}+\delta  f_{\mu\nu}$ (the metric $f_{\mu\nu}$ 
can be varied even though it is non-dynamical,
\blue{in order to obtain identities similar to the Bianchi identity}). 
 The potential \eqref{pot} then receives the variation   
 \be                \label{dU}
 \delta\,\left({U}\sqrt{-g}\right)=\frac12\left(\dg\, {X}^{\mu\nu}\, 
\delta g_{\mu\nu}-\df\, {Y}^{\mu\nu}\, \delta f_{\mu\nu}\right),
 \ee
 where ${X}^{\mu\nu}={X}^\mu_{~\alpha} g^{\alpha\nu}$ and ${Y}^{\mu\nu}
={Y}_{~\alpha}^{\mu} f^{\alpha\nu}$
 with 
 \bea                   \label{XY}
 X^\mu_{~\nu}&=&\frac{1}{2n^2}\,\dr^\al\left(
 n\, \Sigma^\mu_{~\nu}
 -\frac{\al+1}{2}\,\Sigma^\alpha_{~\alpha}\,\delta^\mu_{~\nu}
 \right),             \nonumber \\
 Y^\mu_{~\nu}&=&\frac{1}{2n^2}\,\dr^{\al+1}\left(
 n\, \Sigma^\mu_{~\nu}
 -\frac{\al}{2}\,\Sigma^\alpha_{~\alpha}\,\delta^\mu_{~\nu}
 \right),
 \eea
 where 
 $
 \Sigma^\mu_{~\nu}=(\hat{S}^n)^\mu_{~\nu}\,.
 $
Consider an infinitesimal diffeomorphism generated by a vector field $\xi^\mu$.
It induces the variations of both metrics, 
 \be  
  \delta g_{\mu\nu}=\overset{(g)}{\nabla}_{(\mu} \overset{(g)}{\xi}_{\nu)},~~~~~~~~~~~
 \delta f_{\mu\nu}=\overset{(f)}{\nabla}_{(\mu} \overset{(f)}{\xi}_{\nu)},
 \ee
 where $\overset{(g)}{\xi}_\mu=g_{\mu\sigma}\,\xi^\sigma$ and 
 $\overset{(f)}{\xi}_\mu=f_{\mu\sigma}\,\xi^\sigma$ while $\overset{(g)}{\nabla}$ and 
 $\overset{(f)}{\nabla}$ are the covariant derivatives with respect to the g-metric and 
f-metric, respectively. 
 Inserting this to \eqref{dU}, integrating over the manifold, dropping the boundary term and using 
 the fact that $U$ is a scalar and hence its integral 
does not change under diffeomorphisms, gives the identity 
 \be                           \label{id}
 \dg\, \overset{(g)}{\nabla}_\mu X^\mu_{~\nu}=\df\, \overset{(f)}{\nabla}_\mu Y^\mu_{~\nu}\,.
 \ee
 Let us now vary the whole action \blue{only} with respect to $g_{\mu\nu}$. 
 Setting the variation to zero, yields the equations 
  \be                       \label{eq}
 G^\mu_{~\nu}+m^2\,\lambda_g\,\delta^\mu_{~\nu}=m^2 X^\mu_{~\nu}+
 T^{\rm (m)\mu}_{~~~~~\nu}\,,
 \ee
where 
  $T^{\rm (m)}_{~\mu\nu}$ is obtained by varying ${\cal L}_{\rm matter}$. 
If the latter is given by   \eqref{mat} then 
\be
T^{\rm (m)\mu}_{~~~~~\nu}=-\bm{\Lambda}\,\delta^\mu_{~\nu}.
\ee
  %$\delta{\cal L}_{\rm mat}=\gamma T^{\rm mat}_{~\mu\nu} \delta g^{\mu\nu}$

\subsection{Subsidiary conditions}

In view of  the Bianchi identities 
$\overset{(g)}{\nabla}_\mu G^{\mu}_{~\nu}=0$ 
and owing to the matter conservation condition 
$\overset{(g)}{\nabla}_\mu T^{{\rm (m)}\mu}_{~~~~~\nu}=0$,
equations \eqref{eq} imply that 
\be                       \label{consX}
\overset{(g)}{\nabla}_\mu X^\mu_{~\nu}=0. 
\ee
This in turn implies, in view of the identity \eqref{id}, that 
\be                       \label{consY}
\overset{(f)}{\nabla}_\mu Y^\mu_{~\nu}=0.
\ee
Now, according to \eqref{Shh}, one has $S^\mu_{~\nu}=\delta^\mu_{~\nu}+\chi^\mu_{~\nu}$,
where $\chi^\mu_{~\nu}$ vanishes if $g_{\mu\nu}=f_{\mu\nu}$.
The weighted power of this matrix can be represented similarly, 
\be                                \label{prim}
\dr^{\al+1}\,(S^n)^\mu_{~\nu}=
\delta^\mu_{~\nu}+\Psi^\mu_{~\nu},
\ee
where $\Psi^\mu_{~\nu}$ vanishes when the two metrics coincide. 
This relation is in fact equivalent to the one in \eqref{g-g1},
assuming that the indices are moved with the f-metric so that 
\be
((\hat{g}^{-1})^n)^{\mu\nu}=(\hat{S}^n)^\mu_{~\sigma}\,f^{\sigma\nu},~~~~~~
\hh^{\mu\nu}=\hh^\mu_{~\sigma}\,f^{\sigma\nu}. 
\ee
Using \eqref{prim} reduces \eqref{consY}  to 
\be                     \label{psi0}
\overset{(f)}{\nabla}_\mu\left( \Psi^\mu_{~\nu}
-\frac{\al}{2n}\,(\Psi^\alpha_{~\alpha})\,\delta^\mu_{~\nu}\right)=0. 
\ee
If the reference metric is chosen to be flat Minkowski, 
$f_{\mu\nu}=\eta_{\mu\nu}$, then the derivatives $\overset{(f)}{\nabla}_\mu$ become 
ordinary partial derivatives and \eqref{psi0}  reduces to the Lorentz-Hilbert condition 
for the $\Psi$-field, 
\be                   \label{psi}
\partial_\mu \Psi^\mu_{~\nu}+q\,\partial_\nu \Psi^\alpha_{~\alpha}=0, 
\ee
with $q=-\al/(2n)$. This explains the OP's trick  -- the linear in 
$ \Psi^\mu_{~\nu}$ subsidiary condition  \eqref{psi} 
indeed follows from the non-linear field equations \eqref{eq}. 
This explains also why this condition is not manifestly covariant -- the field equations 
 \eqref{eq} are covariant if only both metrics are allowed to simultaneously transform, 
 but the covariance is lost as soon as the metric $f_{\mu\nu}$ is fixed (unitary gauge). 

Even though the OP theory is formulated entirely in terms of the metrics $g_{\mu\nu}$ 
and $f_{\mu\nu}$, according to the OP's philosophy, the metric $g_{\mu\nu}$
is only a secondary object. The primary object is supposed to be the graviton field 
$ \Psi^\mu_{~\nu}$ determining  the metric $g_{\mu\nu}$ via \eqref{prim}. 
Setting the philosophy aside, the mathematical statement is that in the OP theory 
there exists the {invertible} non-linear transformation \eqref{prim} 
expressing the metric $g_{\mu\nu}$ 
in terms of $\hh^{\mu\nu}$  
such that the condition $\partial_\mu Y^\mu_{~\nu}=0$ becomes {\it linear} in 
$\hh^{\mu\nu}$.

One should stress at the same time that the tensor $Y^\mu_{~\nu}$  can be defined
via \eqref{dU} in any bimetric theory,
for any choice of the mass term $U$. 
The  condition \eqref{consY} will always hold on-shell, and 
setting 
$f_{\mu\nu}=\eta_{\mu\nu}$ one always obtains  $\partial_\mu Y^\mu_{~\nu}=0$. 
However,  $Y^\mu_{~\nu}$ will in general contain non-linear terms that cannot 
 be absorbed by redefining the variables \blue{via \eqref{prim}}, hence   conditions 
$\partial_\mu Y^\mu_{~\nu}=0$  will not 
have the Lorentz-Hilbert form needed for  the ``spin limitation". 

For example, in  
the dRGT theory with the potential \eqref{dRGT}
one has 
\be
{Y}^\mu_{~\nu}=-\frac{1}{\det(\hat{\gamma } )  }\left\{
\left(\beta_1+\beta_2\, U_1+\beta_3\,U_2    \right)
{{\gamma}}^\mu_{~\nu}-(\beta_2+\beta_3\, U_1)\,(\hat{\gamma}^2)^\mu_{~\nu}
+\beta_3\, (\hat{\gamma}^3)^\mu_{~\nu}
\right\}.
\ee
If $\beta_2=\beta_3=0$ then the conservation condition $\partial_\mu Y^\mu_{~\nu}=0$
reduces to 
\be
\partial_\mu\left(\frac{1}{\det(\hat{\gamma } )}\, {{\gamma}}^\mu_{~\nu}\right)=0,
\ee
\blue{which  has the form \eqref{prim} and 
can be linearized by setting}
\be
\frac{1}{\det(\hat{\gamma } )}\, {{\gamma}}^\mu_{~\nu}={\delta}^\mu_{~\nu}+{\hh}^\mu_{~\nu}\,,
\ee
which yields $\partial_\mu{\hh}^\mu_{~\nu}=0$. \blue{However, the same trick does not work 
for generic values of $\beta_k$. }
For example, if $\beta_1=\beta_3=0$ and $\beta_2\neq 0$ then one obtains 
\be
\partial_\mu\left(\frac{1}{\det(\hat{\gamma } )}\,\left([\hat{\gamma}] {{\gamma}}^\mu_{~\nu}
-(\hat{\gamma}^2)^\mu_{~\nu}\right)\right)=0, 
\ee
\blue{
which cannot be linearized by applying \eqref{prim}. Of course, this can be linearized 
by a different transformation. However, within the OP approach, the transformation should be 
the same as the one OP used to derive the Einstein-Hilbert kinetic term starting from the 
linear theory, hence it must have the form \eqref{prim}, which is equivalent to \eqref{g-g1}. 
}

Summarizing, the OP  potential is adjusted in such a way that the 
tensor  $Y^\mu_{~\nu}$ has the structure 
$Y^\mu_{~\nu}=D^\mu_{~\nu}+const.\times  [\hat{D}]\,\delta^\mu_{~\nu}$  
where $D^\mu_{~\nu}$ is the weighted power of $g_{\mu\nu}$. 
Changing the variables via
$D^\mu_{~\nu}=\delta ^\mu_{~\nu}+\hh^\mu_{~\nu}$,  the one-shell condition 
$\partial_\mu Y^\mu_{~\nu}=0$ assumes the linear form 
$\partial_\mu \Psi^\mu_{~\nu}+q\,\partial_\nu \Psi^\alpha_{~\alpha}=0$. 
 According to OP, this property is very important, since the linear 
condition for $\hh^{\mu\nu}$ restricts the spin of states 
in the theory. \blue{To understand what this property gives in practical terms, 
we shall now study the phenomenology of the theory. }

 \section{Solutions with proportional metrics}
 \setcounter{equation}{0}
We shall first consider the simplest solutions for which the reference metric $f_{\mu\nu}$ 
is not fixed once and forever 
but related to the physical metric via 
\be                   \label{prop}
f_{\mu\nu}=\xi^2\, g_{\mu\nu}
\ee
with constant $\xi$. This implies that 
\be                                    \label{XX}
S^\mu_{~\nu}=\xi^2 \delta^\mu_{~\nu}~~~~~\Rightarrow~~~~~~
X^\mu_{~\nu}=\lambda_g\,\xi^{2(n-2s)} \delta^\mu_{~\nu}\,,
\ee
and the field equations \eqref{eq} reduce to 
\be                         \label{ee}
G_{\mu\nu}+\Lambda\,g_{\mu\nu}=0
\ee
with
\be                                           \label{Lam}
\Lambda=\bm{\Lambda}+m^2\lambda_g\left[1-\xi^{2(n-2s)  }   \right].
\ee
It follows that the solution is an Einstein space with $R_{\mu\nu}=\Lambda\, g_{\mu\nu}$,
for example the de Sitter space. 
If the parameters are adjusted such that $\Lambda=0$, 
then the flat space will be a solution. We emphasize once again that to 
different solutions there correspond different reference metrics defined via \eqref{prop}.

Let us now analyze the stability of such solutions. To this end, we consider 
small perturbations of the g-metric without changing the reference metric, 
\be
g_{\mu\nu}\to g_{\mu\nu}+\h_{\mu\nu},~~~~~~~~f_{\mu\nu}\to f_{\mu\nu}, 
\ee
hence 
\be
g^{\mu\nu}\to g^{\mu\nu}-\h^{\mu\nu}+\ldots \,, 
\ee
with 
$\h^{\mu\nu}=g^{\mu\alpha}g^{\nu\beta} \h_{\alpha\beta}$. 
Let us consider the linear in perturbations part of the field equations, 
\be                      \label{pert}
\delta G_{\mu\nu}+(\bm{\Lambda}+m^2\lambda_g)\,\delta g_{\mu\nu}=m^2 \delta X_{\mu\nu}\,,
\ee
where
\be
\delta X_{\mu\nu}=\delta\left( g_{\mu\sigma}X^\sigma_{~\nu}  \right)=h_{\mu\sigma}\,X^\sigma_{~\nu}
+g_{\mu\sigma}\,\delta X^\sigma_{~\nu}=\lambda_g\,\xi^{2(n-2s)}\,h_{\mu\nu}+g_{\mu\sigma}\,\delta X^\sigma_{~\nu}.
\ee
To calculate the variation $\delta X^\sigma_{~\nu}$, we notice that the tensor $X^\mu_{~\nu}$ defined by \eqref{XY} 
contains  powers of the tensor $S^\mu_{~\nu}=g^{\mu\sigma}f_{\sigma\nu}$,  whose variation is 
\be
\delta S^\mu_{~\nu}=\delta g^{\mu\sigma} f_{\sigma\nu}=-h^{\mu\sigma}f_{\sigma\nu}
=-\h^\mu_{~\sigma}S^\sigma_{~\nu}=
-\xi^2 \h^\mu_{~\nu}\,.
\ee
The background tensor  $S^\mu_{~\nu}=\xi^2\delta^\mu_\nu$ is proportional to the unit tensor,
hence  it 
commutes with $\delta S^\mu_{~\nu}$, which implies that  
\be
\delta{\Sigma}^\mu_{~\nu}=\delta ({S}^n)^\mu_{~\nu}
=n\,\delta {S}^\mu_{~\sigma}\, (S^{n-1})^\sigma_{~\nu}=-n{h}^\mu_{~\sigma}\,(S^n)^\sigma_{~\nu}
=-n\xi^{2n}\,h^\mu_{~\nu}. 
\ee
It follows also that 
\be
\delta(\det\hat{S})=\delta (\det{\hat{g}}^{-1})\det\hat{f}=-\det(\hat{S})\, h
\ee
with $h=h^\mu_{~\mu}$ 
and hence 
\be
\delta(\phi^s)=\delta\left(\det \hat{S})^{-s/2}\right)=\frac{s}{2}\,\phi^s\,h=\frac{s}{2}\,\xi^{-4s}h.
\ee
As a result, 
\bea
\delta X^\mu_{~\nu}
&=&\frac{s}{2}\,h\,X^\mu_{~\nu}-\frac{1}{2}\,\xi^{2n-4s}
\left(n h^\mu_{~\nu}-\frac{s+1}{2}\,h\delta^\mu_{~\nu}\right) \nn \\
&=&\xi^{2n-4s}\left(
-\frac12 \,h^\mu_{~\nu}+\left[\frac{s}{2}\,\lambda_g+\frac{s+1}{4n}\right]h\delta^\mu_{~\nu}
\right),
\eea
and the perturbation equations \eqref{pert} assume the form 
\be                                                        \label{HIG}
E_{\mu\nu}\equiv \delta G_{\mu\nu}+\Lambda\,h_{\mu\nu}+\frac{M^2}{2}\,
(h_{\mu\nu}-\lambda h\, g_{\mu\nu})=0
\ee
with $\Lambda$ defined by \eqref{Lam} and with 
\be
M^2=m^2 \xi^{2n-4s}
\ee
while 
 \be             \label{zeta}
 \lambda=1+\frac{1-(n-2\al-1)^2-3n^2}{4n^2}\equiv 1+\zeta. 
 \ee
For $\lambda=1$ Eqs.\eqref{HIG} reduce to those studied by Higuchi 
to describe massive gravitons in de Sitter space \cite{Higuchi}, 
the parameter $M$ then determines the 
graviton mass. The equations show the FP property in this case -- 
they propagate only 5 DoF.
However,  for $\lambda\neq 1$ the number
of DoF is 6. Let us remind the corresponding counting  argument. 

There are 10 equations in \eqref{HIG}, where one has 
 \bea
\delta G_{\mu\nu}=&&\left.\frac12\left(
\nabla^\sigma \nabla_\mu \h_{\sigma\nu}
+\nabla^\sigma \nabla_\nu \h_{\sigma\mu}-\Box \h_{\mu\nu}-\nabla_\mu \nabla_\nu h-R\,\h_{\mu\nu}\right)\right.
\nonumber \\
&&+\left.\frac12 \,
g_{\mu\nu}\left( \Box h-\nabla^\alpha\nabla^\beta h_{\alpha\beta} +R^{\alpha\beta}h_{\alpha\beta} \right)\right.. 
\eea
For $R_{\mu\nu}=\Lambda\,g_{\mu\nu}$ and $R=4\Lambda$ there is the 
identity relation
$
\nabla^\mu (\delta G_{\mu\nu}+\Lambda h_{\mu\nu})=0, 
$
hence  taking the divergence of \eqref{HIG}  yields four constraints 
 \be                 \label{FP1}
\frac{M^2}{2}\left(\nabla^\mu \h_{\mu\nu}-\lambda \nabla_\nu h\right)=0
 \ee
from \blue{which} $\nabla^\mu \h_{\mu\nu}=\lambda \nabla_\nu h$. Using these relations, 
Eqs.\eqref{HIG} reduce to 
\bea
-\Box h_{\mu\nu}+(2\lambda-1)\nabla_{\mu\nu}h
-2R_{\mu\alpha\nu\beta}\,h^{\alpha\beta} 
&+&\left[(1-\lambda)\Box h+\Lambda h  \right] g_{\mu\nu} \nonumber \\
&+&M^2\left[h_{\mu\nu}-\lambda h\, g_{\mu\nu}  \right]=0,
\eea
and taking the trace one obtains 
\be              \label{FP2}
2(1-\lambda)\Box h+\left[2\Lambda+(1-4\lambda)M^2  \right]h=0. 
\ee
If $\lambda=1$ then this yields 
\be 
\left(2\Lambda-3M^2  \right)h=0, 
\ee
implying the fifth constraint, $h=0$ (unless in the partially massless limit $2\Lambda=3M^2$). 
Therefore, the number of DoF is the number of components of $h_{\mu\nu}$ 
minus the number of constraints in \eqref{FP1},\eqref{FP2}, which gives 
$10-5=5$. This corresponds to the Fierz-Pauli theory.

If $\lambda\neq 1$ then the FP property is lost, because 
there is the non-trivial kinetic term in \eqref{FP2}, 
hence the trace $h$ becomes a dynamical mode, so that 
there are 6 DoF. 

 %\iffalse  
\begin{figure}[th]
\hbox to \linewidth{ \hss
%	\rotatebox{-90}\resizebox{14cm}{12cm}{\includegraphics{f.eps}}
	\resizebox{12cm}{9cm}{\includegraphics{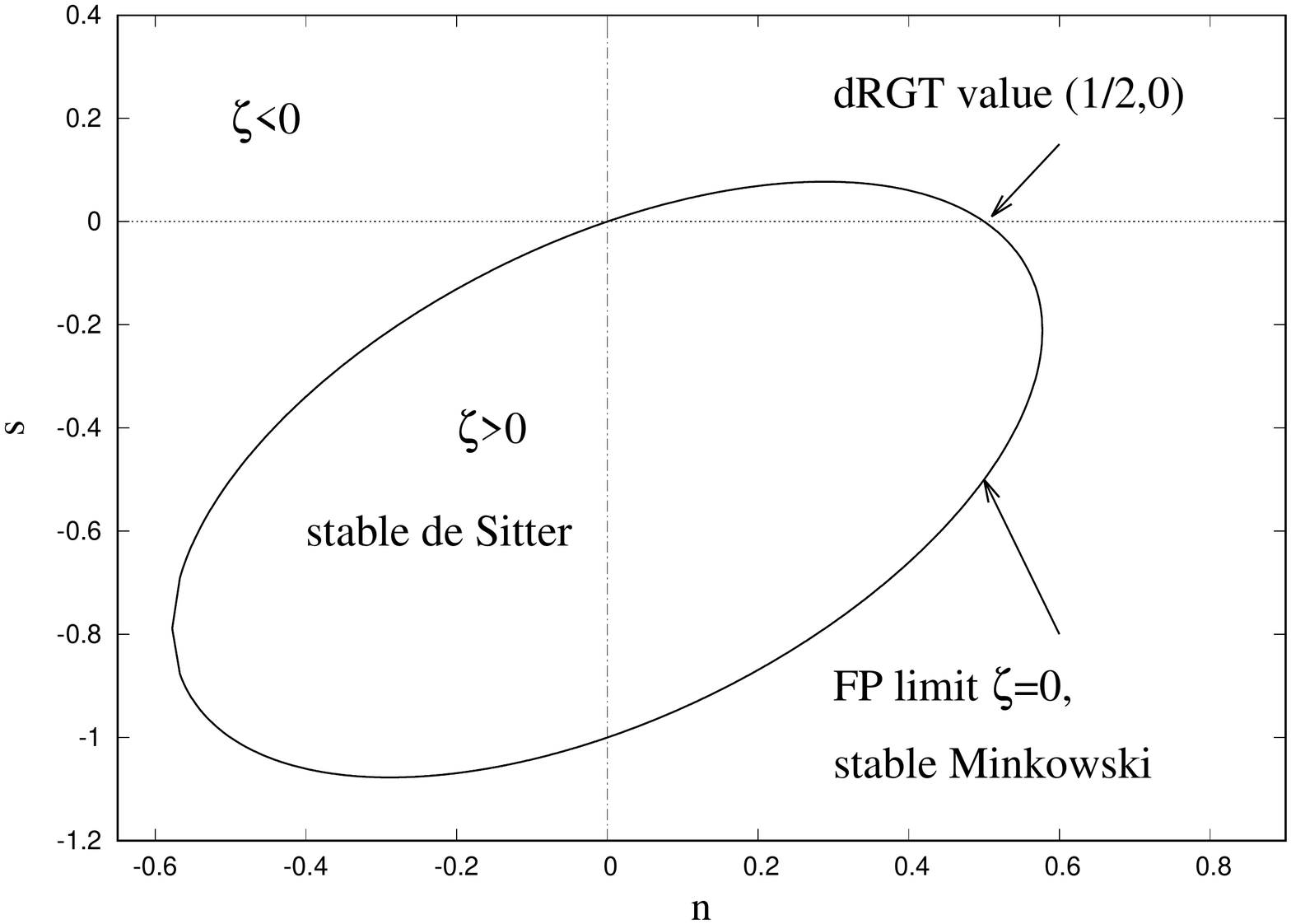}}
%	\resizebox{14cm}{12cm}{\includegraphics{f.eps}}
%\includegraphics[width=1.3\textwidth, angle =90 ]{f.eps}
\hss}
\caption{The FP subset of the OP theory.}
%\caption{Solution of Eqs.\eqref{eeqs} for the initial values \eqref{in1}--\eqref{in3}. }
\label{Fig1}
\end{figure}
%\fi

%\begin{figure}[!btp]\begin{center}
%\rotatebox{-1}{\includegraphics[width=0.9\textheight,angle=-89]{f.eps}}
%\caption{Modulator's block diagram}\label{blockdiagram}
%\end{center}\end{figure}

Therefore, the OP theory respects
 the FP property if the parameter $\lambda$ 
defined by  \eqref{zeta} is equal to one, hence if
\be
(n-2\al-1)^2+3n^2=1. 
\ee
This   defines an ellipse in the $n\al$-plane shown in Fig.\ref{Fig1}. 
Points of this ellipse correspond to the special case of the  OP theory in which the FP property is respected 
and there are only 5 DoF
around the de Sitter (or flat for $\Lambda=0$) space. 
Points not belonging  to the ellipse correspond to the generic OP 
theory with  6 DoF.

If $\Lambda=0$ and the background geometry is flat, then 
 among the 6 DoF there is ghost -- 
a mode with a negative kinetic energy. 
One might think that all of the OP theories with 6 DoF  are unphysical, since flat space is  unstable
in such theories. 
However, as we shall now see,
the whole interior of the ellipse corresponds to theories in which the 
de Sitter space is stable \blue{at the level of liner perturbations}. 

 \section{Stability conditions}
 \setcounter{equation}{0}
 Let us assume the background metric $g_{\mu\nu}$
  to be of the Friedmann-Lema$\hat{\i}$tre-Robertson-Walker (FLRW) type, 
 \be
 ds_g^2=-N^2(t)\,dt^2+a^2(t)\, \delta_{ij}\, dx^i dx^j=\frac{1}{\xi^2}\, ds_f^2, 
 \ee
 which fulfills background equations \eqref{ee} with $\Lambda\equiv 3H^2>0$, 
 \be
 \frac{\dot{a}^2}{N^2 a^2}=H^2.  
 \ee
 This describes  the de Sitter space expressed in the spatially flat slicing. 

Perturbing the solution, 
 \be                     \label{pert0}
 g_{\mu\nu}\to g_{\mu\nu}+h_{\mu\nu},~~~~~~~f_{\mu\nu}\to f_{\mu\nu},
 \ee
the perturbations can be decomposed into the 
 scalar, vector, and tensor parts via 
 \bea      \label{h}
 h_{00}&=&-N^2 {\bf S}_3,   \nn \\
 h_{0i}&=&Na \left( \partial_i {\bf S}_4+{\bf W}_i  \right),   \nonumber \\
 h_{ik}&=&a^2\left(  
 {\bf S}_1\,\delta_{ik}+\partial^2_{ik}{\bf S}_2+\partial_i {\bf V}_k+\partial_k {\bf V}_i+{\bf D} _{ik}
 \right),
 \eea
 where 
 \be
\sum_k \partial_k {\bf V}_k=\sum_k \partial_k {\bf W}_k=0,~~~~~
\sum_k \partial_k {\bf D}_{ki}=0,~~~~\sum_k {\bf D}_{kk}=0.
 \ee
 The spatial dependence of the modes is given by the plane waves
$\exp(i\,{\bf px})=\exp(ipz)$,  where the wave vector 
 can be oriented along the 3-rd (z) axis. 
The amplitudes ${\bf S}_4$ and ${\bf V}_k$ have dimension of length,
while ${\bf S}_2$ has dimension of length squared. To pass to dimensionless quantities, 
we introduce a mass scale $\mu$ and set 
 \be
 {\bf S}_1=S_1(t) e^{ipz},~~~~~
{\bf S}_2=\frac{1}{\mu^2}\,S_2(t) e^{ipz},~~~~~
{\bf S}_3=S_3(t) e^{ipz},~~~~~
{\bf S}_4=\frac{1}{\mu}\,S_4(t) e^{ipz},~~~~~
 \ee
 the vector amplitudes are chosen as 
 \be
 {\bf V}_k=\frac{1}{\mu}\,[V_{1}(t),V_{2}(t),0]\,e^{ipz},~~~~~~~
 {\bf W}_k=[W_{1}(t),W_{2}(t),0]\,e^{ipz},
 \ee
 while for the tensor modes the only non-trivial components of ${\bf D}_{ik}$ are 
 \be
 {\bf D}_{11}=-{\bf D}_{22}=D_{1}(t)\,e^{ipz},~~~~~
 {\bf D}_{12}={\bf D}_{21}=D_{2}(t)\,e^{ipz}. 
 \ee
The mass scale $\mu$ can be, for example,  the Planck mass $M_{\rm Pl}$, or 
the Hubble rate $H$, or 
the graviton mass $M$. 
However, we prefer not to specify it to be able to consider the limits such as 
 $H\to 0$
or $M\to 0$.

 Inserting everything into the perturbation equations $E_{ik}=0$ \eqref{HIG}, they
 split into three independent groups for the scalar, vector, and tensor modes. These equations 
determine the effective action, which is the sum of three independent terms, 
 \be                  \label{act0}
 I=I_{\rm T}+I_{\rm V}+I_{\rm S}=\frac{M_{\rm Pl}^2 }{4} \int Na^3\, \bar{h}^{\mu\nu}E_{\mu\nu}\, dt\,d^3x,
 \ee
 where the bar denotes complex conjugation. 
 One obtains in the tensor sector 
 \be                 \label{IT}
 I_{\rm T}=\frac{M_{\rm Pl}^2 }{4}\, \int Na^3\left(  
 \frac{1}{N^2}\left(\dot{D}_{1}^2+\dot{D}_{2}^2\right)-\left[M^2+\frac{p^2}{a^2}  \right](D_{1}^2+D_{2}^2)
   \right)dt\,d^3x.
 \ee
 Inspecting the equations in the vector sector one finds that the two amplitudes $W_1$ and $W_2$ can be 
 expressed in therms of $V_1$ and $V_2$, the latter being governed by the action 
 \be                     \label{IV}
 I_{\rm V}=\frac{M_{\rm Pl}^2 }{4\mu^2}\,M^2 \int Na^3\left(  
 \frac{p^2}{N^2(p^2/a^2+M^2)}\left(\dot{V}_{1}^2+\dot{V}_{2}^2\right)-p^2(V_{1}^2+V_{2}^2)
   \right)dt\,d^3x. 
 \ee
 Neither tensor nor vector modes are sensitive to the value of the parameter $\zeta$ 
 describing the deviation from the FP limit. The vector modes become non-dynamical when the mass $M$ 
 tends to zero. 
 
 In the scalar sector, the amplitudes $S_3$ and $S_4$ can be expressed in terms of $S_1$ and $S_2$,
 the effective action for the latter being 
 \be                           \label{IS}
 I_{\rm S}^{(\zeta)}=\frac{M_{\rm Pl}^2 }{4}\, M^2 \int Na^3\sum_{a,b=1,2} \left(  
 \frac{1}{N^2}\,K_{ab}\,\dot{S}_a\dot{S}_b
 +\frac{\cal Q}{N}\,\epsilon_{ab}\,\dot{S}_a S_b - U_{ab}\,S_a S_b
   \right)dt\, d^3 x.
 \ee
 Here the kinetic matrix has components 
 \be                 \label{KK}
 K_{11}=\zeta\,\frac{3a^2M^2+4p^2}{X},~~~
K_{22}=\frac{2a^2H^2p^4}{\mu^4\,X},~~~~
K_{12}=K_{21}=-\zeta\,\frac{a^2 M^2 p^2}{\mu^2 X},
 \ee 
 with 
 \be                          \label{X} 
 X=8H^2p^2+M^2a^2(6H^2-\zeta M^2),
 \ee
  and  one has  
 \be                              \label{QQ}
 {\cal Q}=-2(1+2\zeta)\,\frac{H p^4}{\mu^2 X},~~~~~~~\epsilon_{ab}=-\epsilon_{ba},~~~\epsilon_{12}=1. 
 \ee
 Components of the potential matrix have more complex structure,
 \bea
 U_{11}=\frac{1}{2a^2X^2}\{-32H^2\, p^6 +4a^2[48H^4-6M^2(2\zeta+5)H^2+\zeta M^4]\,p^4 \nn \\
 +2M^2a^4 [144H^4+6(4\xi^2-11\xi-12)M^2H^2+\zeta(5\zeta+6)M^4]\,p^2   \nn \\
 +3M^4a^6(6H^2-\zeta M^2)(6H^2-(3+4\zeta)M^2)\},  \nn \\
 U_{22}=\frac{p^4}{2\mu^4X^2}\{-64\zeta H^4\, p^4 
 -8a^2H^2M^2[(2\zeta-10)H^2+(1+2\zeta-\zeta^2)M^2]\,p^2   \nn \\
 +M^4a^4(6H^2-(1+2\xi)M^2)(6H^2-\zeta M^2)
 \},  \nn  \\
U_{12}=U_{21}=\frac{p^2}{2\mu^2X^2}\{16H^2[ (6\zeta-1)H^2+(1+\zeta)M^2]\,p^4   \nn \\
+2a^2M^2[6(2\zeta-7)H^4-(10\zeta^2-17\zeta-18)M^2H^2-\zeta(1+\zeta)M^4  ]\,p^2  \nn \\
-M^4a^4(6H^2-\zeta M^2)(6H^2-(3+4\zeta)M^2)
\}.
 \eea
We notice that the scalars become non-dynamical in the $M\to 0$ limit.

 Let us analyze the positivity of the kinetic matrix $K_{ab}$. This eigenvalues 
 $\lambda_1$ and $\lambda_2$  of this 
 matrix will be positive definite if the trace  ${\rm tr}( K_{ab})=\lambda_1+\lambda_2$
 and the  determinant $\det (K_{ab})=\lambda_1\lambda_2$ are both positive. One has 
 \bea                       \label{XK}
 \mbox{tr}(K_{ab})&=&\frac{1}{\mu^4\,X}\,(2H^2a^2 p^4+4\zeta \mu^4\,p^2+3\zeta a^2 \mu^4 M^2)
 ~~~\stackrel{p\to\infty}{\longrightarrow}
 ~~~\frac{a^2 p^2}{4\mu^4}  +{\cal O}(1),     \nn  \\
 \det(K_{ab})&=&\frac{1}{\mu^4 X}\,\zeta a^2 p^4
 ~~~\stackrel{p\to\infty}{\longrightarrow}
 ~~~\frac{\zeta a^2}{8\mu^4H^2} \,p^2+{\cal O}(1). 
 \eea
 Therefore, as long as $\zeta>0$, 
 the kinetic matrix is always positive-definite  in the UV limit 
 $p\to\infty$, and it will be positive definite for {\it any} momenta  if 
\be
6H^2>\zeta M^2
\ee
since $X>0$ in this case. 
 This conclusion applies only when the Hubble rate of the background 
 metric  is non-zero, $H\neq 0$. 

  If $H=0$ then the background metric 
 is flat and one can set $a=1$, hence \eqref{X} yields $X=-\zeta M^4$. 
Eq.\eqref{XK} then gives 
 \be                            \label{Mink}
 \mbox{tr}(K_{ab})=-\frac{4p^2+3M^2 }{M^4}<0,~~~~~~
 \det(K_{ab})=-\frac{p^4}{\mu^4 M^4}<0,
 \ee
 hence one of the two eigenvalues $\lambda_1$ and $\lambda_2$ is always 
 negative. 
 
 As a result,  there is  ghost around  flat space if $\zeta\neq 0$. This fact is of course well-known. 
 What is new is that  the ghost becomes a benign mode in the de Sitter space 
 if $\zeta>0$. 
 
 One may wonder if there are \blue{gradient instabilities} in the system. 
 The sound speed $C_{\rm S}$ is determined by the algebraic equation 
 \be                           \label{speed}
 \det\left(\frac{p^2C_{\rm S}^2}{a^2}\,K_{ab}+i\,\frac{p\,C_{\rm S}}{a}\,{\cal Q}\,\epsilon_{ab}-U_{ab} \right)=0.
 \ee
 This gives two different values for $C_{\rm S}^2$ determining  the speed of the 
scalar component of the massive graviton and that of the 6-th mode. 
Both  have the same UV limit,
 \be                         \label{speed1}
 \lim_{p\to\infty}C_{\rm S}^2=1. 
 \ee
 Since $C_{\rm S}^2>0$, there are no \blue{gradient instabilities} in this limit. 
This conclusion applies 
both for the de Sitter space ($H\neq 0$) and for Minkowski space ($H=0$).

  \subsection{The FP limit}
  
 Let us finally see what happens when $\zeta\to 0$.  The only 
 non-vanishing component of the kinetic matrix $K_{ab}$ in this limit is $K_{22}$,
 hence the amplitude $S_1$ becomes non-dynamical and can be algebraically 
 expressed in terms of $S_2$.  Injecting this expression back to the action yields 
  \be
 I_{\rm S}^{(\zeta=0)}=\frac{3M_{\rm Pl}^2}{4\mu^4}\,M^2(M^2-2H^2)\int Na^3 \left(  
 \frac{1}{N^2}\,K\dot{S}_2^2-US_2^2
   \right)\,dt\, d^3x
 \ee
 with 
 \be
 K=\frac{a^4p^4}{Y},~~~~U=a^2p^4\,
 \frac{16H^2a^2p^4+(p^2+a^2M^2)Y}{Y^2},
 \ee
where $Y=4p^4+3a^2(M^2-2H^2)(4p^2+3M^2a^2)$. We see that the scalar mode becomes 
non-dynamical either for $M=0$ (massless limit) or for $M^2=2H^2$  (partially 
massless limit). The kinetic therm is always positive if $M^2>2H^2$ but for $M^2<2H^2$
it becomes negative and the scalar mode becomes the 
(Higuchi) ghost. The speed of sound is equal to one in the UV limit. 
Sending $H\to 0$, one can see that the flat space is stable in this case. 

\subsection{Generic massive gravity }

The above results are actually quite general and 
apply not only in the OP theory but also in the 
generic bimetric theory \eqref{act}. 
Specifically, introducing 
$
H^\mu_{~\nu}=\delta^\mu_{~\nu}-S^\mu_{~\nu}
$
with $S^\mu_{~\nu}=g^{\mu\sigma}f_{\sigma\nu}$, 
the potential $U$ in \eqref{act} can be expanded as
\be
U=a_0+a_1[\hat{H}]+a_2[\hat{H}^2]+a_3[\hat{H}]^2+\ldots 
\ee
If the matter 
term in \eqref{act} is given by \eqref{mat}, then the field equations read 
\be                   \label{dSE}
G_{\mu\nu}+{\bm\Lambda}g_{\mu\nu}+m^2(2a_1 f_{\mu\nu}+4a_2 f_{\mu\sigma}H^\sigma_{~\nu}
+4a_3[\hat{H}]f_{\mu\nu}+(a_0+a_1[\hat{H}]) g_{\mu\nu}+\ldots)=0. 
\ee
If $f_{\mu\nu}=g_{\mu\nu}$ then $H^\mu_{~\nu}=0$ and the equations reduce to 
\be
G_{\mu\nu}+\Lambda g_{\mu\nu}=0~~~~~~{\rm with}~~~~~~
\Lambda={\bf \Lambda}+m^2(a_0+2 a_1).
\ee
Therefore, if the reference metric is chosen to be de Sitter,
then the theory admits a solution for which the physical metric is also de Sitter. 
Consider perturbations
$
g_{\mu\nu}=f_{\mu\nu}\to f_{\mu\nu}+h_{\mu\nu}
$
with fixed  $f_{\mu\nu}$. Linearizing Eqs.\eqref{dSE} with respect to 
$h_{\mu\nu}$ then yields precisely the Higuchi equations \eqref{HIG} with 
\be
M^2=8 m^2 a_2,~~~~~~\lambda=1+\zeta=-\frac{a_3}{a_2}. 
\ee
Therefore, the above analysis directly applies  and one can say at once that the 
flat space (obtained if $\Lambda=0$) is stable 
if $a_2+a_3=0$, 
while the de Sitter space is stable if $\lambda>1$.

\section{More general cosmologies}
\setcounter{equation}{0}

Let us now study more general solutions of the OP theory.
We shall be considering FLRW cosmologies described by 
\bea                \label{gf}
ds_g^2&=&-N^2(t)\,dt^2+a^2(t)\,\Omega_{ij}dx^i dx^j,  \nonumber \\
ds_f^2&=&-N_f^2(t)\,dt^2+a_f^2(t)\, \Omega_{ij}dx^i dx^j, 
\eea
where $\Omega_{ij}dx^i dx^j$ is the metric of the maximally symmetric 3-space 
with constant curvature $K=0,\pm 1$. We can assume without loss of generality that the 
functions $N,N_f,a,a_f$ are positive. The time reparameterization freedom 
can be used to impose one gauge condition,  for example, one can set $N=1$, or $N_f=1$, or $a=t$. 
Denoting 
\be                \label{xi}
\frac{a_f}{a}\equiv \xi,~~~~~~~
\frac{N_f}{N}\equiv \y\,\xi\,,
\ee
and injecting everything to \eqref{Sh} and \eqref{1} yields 
\be
S^\alpha_{~\beta}=\xi^2\,{\rm diag}[\,\y^2,1,1,1],~~~~~~~~~
\phi=(\y\xii^4)^{-1}.
\ee
The function  $c$ is the speed of light 
measured with respect to the reference metric. 

\subsection{Stability conditions }

Before we study solutions  of the form \eqref{gf},
let us describe their stability conditions. 
These conditions are derived in the Appendix\footnote{\blue{Eqs. \eqref{zzz} and \eqref{ccc0} are 
derived in the Appendix for the spatially flat $K=0$ background, but they are 
valid also for the spatially open $K=1$ and closed $K=-1$ backgrounds.} },
without imposing the background field equations
and only assuming that  $\dot{a}\neq 0$. 
It turns out that backgrounds 
\eqref{gf} may in general accommodate 
ghost and \blue{gradient instabilities}. Ghosts (excitations with a negative kinetic energy)
will be absent in the UV limit, for momenta 
much larger than the Hubble parameter,
\be                                     \label{UV}
p\gg H=\frac{\dot{a}}{Na},
\ee
if the following condition holds:
\be                  \label{zzz}
 \zeta(\y)\equiv -\frac{(2n-s-1)(2n-s)\y^{2n}+3s(s+1) }{4n^2}>0. 
 \ee 
However, soft ghosts with a wavelength 
of the order or larger than the cosmological horizon may still be present. 
Notice that if 
$c\to 1$ then  $\zeta(c)$ reduces to 
$\zeta$ defined by \eqref{zeta}. 

\blue{The gradient instability} is  
characterized by an imaginary 
sound speed. The sound speed $C_{\rm S}$ is determined by the 
algebraic equation (again assuming the UV limit \eqref{UV})
\be                                               \label{ccc0}
 2C^2_{\rm V}\,\zeta(\y)\,C_{\rm S}^4
+\left[C^4_{\rm V}\left(4Z\,\zeta(\y)-\omega^2 \right)+\y^2\right] C_{\rm S}^2
 +2\y^2C^2_{\rm V}Z=0,
 \ee
where
\bea
 \omega&=&-\frac{1+\y^2}{2C^2_{\rm V}}+\frac{2s+1}{2n}\,(1+\y^{2n})
-\frac{s(s+1)}{2n^2}\,(3+\y^{2n}),   \nonumber \\
 Z&=&-\frac{1}{4n^2}\left[2n(2n-2s-1)+s(s+1)(3+\y^{2n})  \right], 
\nonumber \\
 C^2_{\rm V}&=&n\,\frac{\y^2-1}{\y^{2n}-1}. 
 \eea
There are in general two different values of $C_{\rm S}^2$ which fulfill \eqref{ccc0},
hence two different sound speeds. 
\blue{The gradient instability} will be absent if 
\be
C_{\rm S}^2>0. 
\ee
In the $c\to 1$ limit Eq.\eqref{ccc0} reduces to 
$
 2\zeta(C_{\rm S}^2-1)^2=0,
$
 in agreement with the previous result \eqref{speed1}.

 \subsection{Simplest solutions}

Let us now consider the field equations for metrics \eqref{gf}. 
Eq.\eqref{XY} yields  the following non-zero components for the tensor $X^\mu_{~\nu}$:
  \bea               \label{eee}
 X^0_{~0}&=&\frac{1}{2n^2}\, \y^{-s} \xi^{2n-4s}\left(n\, \y^{2n}-\frac{\al+1}{2}\left( \y^{2n}+3\right)\right),  \nonumber \\
 X^1_{~1}=X^2_{~2}=X^3_{~3}&=&
 \frac{1}{2n^2}\, \y^{-s} \xi^{2n-4s}\left(n-\frac{\al+1}{2}\left(\y^{2n}+3\right)
 \right). 
 \eea
The field equations \eqref{eq}  then reduce to
 \bea                                \label{eqn}
 \frac{3\dot{a}^2 }{N^2 a^2}+\frac{3K}{a^2}&=&m^2\lambda_g+\bm{\Lambda}-m^2 X^0_{~0}\,,  \\
\frac{1}{N}\,\left(\frac{\dot{a}}{Na} \right)^{\bm\cdot}-\frac{K}{a^2}&=&\frac{m^2}{2}(X^0_0-X^1_1).                             \label{eqna}
 \eea
These two equations plus an additional gauge condition 
 do not determine all four functions $N,N_f,a,a_f$  and   one  of them 
remains  free. This is the consequence of the fact 
that the system is undetermined because the reference metric is not yet completely 
specified.  An extra assumption is needed  
 in order to determine all four functions. 

As the simplest option, let us assume the physical metric to be flat. 
There are two possibilities  for this, for  the flat 
metric of the form \eqref{gf} can be either  Minkowski,
\be
\dot{a}=0,~~~~~K=0,
\ee
\blue{or} Milne, 
\be
\dot{a}=N\neq 0,~~~~~K=-1. 
\ee
In both cases Eq.\eqref{eqna} requires that $X^0_0=X^1_1$ which implies that 
$c=1$. Eq.\eqref{eqn} then reduces to 
\be
0=m^2\lambda_g+\bm{\Lambda}-\frac{m^2}{2n^2}\, \left(n-s-1\right)\xi^{2n-4s}\,,
\ee
which determines a constant value for $\xi$, hence the reference metric is also 
either Minkowski or Milne, respectively. 

The Minkowski space has already been 
considered above. Its stability is determined by the arguments 
given around Eq.\eqref{Mink} -- it is stable only in the FP limit corresponding to the 
ellipse in Fig.\ref{Fig1}. 

For the Milne solution one has  $\dot{a}\neq 0$ 
and one can apply Eq.\eqref{zzz}. 
This gives the same result as for the de Sitter space
considered above: it is stable everywhere in the region inside the ellipse
in Fig.\ref{Fig1}, where the Minkowski space is unstable. This sounds odd, since the Milne space 
is merely a sector of  Minkowski space expressed in different coordinates.
If $T,R$ are the Minkowski time and radial coordinate, then one has 
\be
T=a(t)\cosh(r),~~~~R=a(t)\sinh(r),
\ee
hence the Milne coordinates $t,r$ cover the interior of the future light cone. 
How can it be that the unstable Minkowski 
space 
 becomes stable when expressed in different 
coordinates~? The answer is that the stability condition \eqref{zzz}
guarantees the absence of ghosts in the UV limit \eqref{UV}, but there could still be soft ghosts 
with momenta
\be
p\leq H=\frac{\dot{a}}{Na}=\frac{1}{a}=\frac{1}{\sqrt{T^2-R^2}}.
\ee
It follows that, although the Milne space does not have UV ghosts, it must contain 
soft ghosts with wavelengths of the order or larger than the Milne horizon. 
One may then argue that the Minkowski space too is actually unstable only with respect 
to long wave ghosts. At first glance, this 
contradicts  the fact that the ghost modes in Minkowski space exist for any momenta.
However, Eq.\eqref{XK}
shows that the leading UV contributions come from interactions of
perturbations with the background curvature, which 
 suggests   that the kinetic energy of
perturbations around Minkowski should  be dominated  by nonlinear
interactions. Therefore, the linear ghost instability of Minkowski space in
the UV limit could presumably be cured by the nonlinear terms.

\subsection{\blue{Solutions with constant $\xi$}}

To study general  solutions of Eqs.\eqref{eqn}, \eqref{eqna},
 it is convenient to consider their 
consequence: the conservation condition
 \be                            
  \left(X^0_{~0}\right)^{\mbox{.}}=3\,\frac{\dot{a}}{a}\left(X^1_{~1}-X^0_{~0}\right).      \label{eqnn}
 \ee
If $\dot{a}\neq 0$ then this replaces the second order 
equation \eqref{eqna}. This condition 
can be represented in the form 
 \be                                \label{eqn1}
 \frac{\dot{\y}}{\y}+F_1(\y)\,\frac{\dot{a}}{a}+F_2(\y)\,\frac{\dot{\xi}}{\xi}=0,
 \ee
 where
 \be
 F_1(\y)=\frac{6n(\y^{2n}-1)}{W},~~~~F_2(\y)=\frac{2(n-2s)((2n-s-1)\y^{2n}-3(s+1)  ) }{W}, 
 \ee
with  $W=(2n-s-1)(2n-s)\,\y^{2n}+3s(s+1)$.
 
 As a result, the solutions are obtained 
by solving Eqs.\eqref{eqn} and \eqref{eqn1}, supplemented by 
an extra condition to totally specify the system. 
For example, one can assume 
 $\xi$ to be a given function of $a$, then  Eq.\eqref{eqn1} becomes 
\be                                \label{eqn2}
\frac{1}{\y}\,\frac{d\y}{da}+F_1(\y)\,\frac{1}{a}+F_2(\y)\,\frac{{\xi^\prime(a)}}{\xi(a)}=0,
\ee
integrating which yields $\y=\y(a)$. Injecting this to \eqref{eee} will give $X^0_{~0}(a)$, 
which will allow  one to integrate Eq. \eqref{eqn}.

 As the simplest option, one can set 
$\xi(a)=\xi_0=const$, in which case  the spatial parts of the two 
 metrics are proportional with the constant factor $\xi_0$. 
Eq.\eqref{eqn2} then 
 reduces to 
 \be                                \label{eqn22}
\blue{\frac{1}{c}\frac{dc}{da} + F_1(\y)\frac{1}{a}=0.}
 \ee
One can fulfill this by setting 
$
 \y=1$,  then the full 4-metrics are conformally related by the factor 
$\xi_0$. This  corresponds to the solutions with proportional 
 metrics already discussed above. 
 The general solution of \eqref{eqn22} for $c\neq const.$ is 
 \be                       \label{a0}
 a=a_0\,\frac{ \y^{\alpha}}{(\y^{2n}-1)^\beta  }~~~~\mbox{with}~~~~
\alpha=\frac{s(s+1)}{2n},~~~~\beta=\frac{n(2n-2s-1)+2s(s+1)}{6n^2}, 
 \ee
 where $a_0$ is the integration constant. 
Inverting this to obtain $c=c(a)$ and 
  injecting to \eqref{eqn} yields  
  \be                                \label{eqn1a}
 \frac{3}{N^2}\frac{\dot{a}^2}{a^2} 
 +\frac{3K}{a^2}=m^2\lambda_g+\bm{\Lambda}
 -\frac{m^2}{4n^2}\, \xi_0^{2n-4s}  
 \left[(2n-s-1)\, \y^{2n-s}(a)-3(\al+1)\y^{-s}(a)\right].
 \ee
Imposing the $N=1$ gauge,  this equation determines 
 $a(t)$, which specifies all four 
metric amplitudes $N,N_f,a,a_f$.

\subsection{\blue{Solutions with Minkowski fiducial metric}}

Let us apply the \blue{procedure outlined above} to construct all solutions 
in the case where the reference metric is flat Minkowski. 
Therefore, the extra assumption is $\dot{a}_f=K=0$. 
Using the definition of $\xi$ in \eqref{xi}, 
one can represent \eqref{eqn1} as 
 \be
 \frac{\dot{\y}}{\y} +  [F_1(\y)-F_2(\y)]\,
 \frac{\dot{a}}{a}
  +F_2(\y)\,\frac{\dot{a}_f}{a_f}  =0,                  \label{eqn222a}\,
\ee
and if  $\dot{a}_f=0$ this reduces to  
   \bea                              \label{eqn3}
a\,\frac{d\y}{da}=  \frac{2\y (n-2s-2)\left[(2n-s)\y^{2n}-3s\right]}{(2n-s-1)(2n-s)\y^{2n}+3s(s+1)}.
\eea
It is clear that 
\be
\y=c\left(\frac{a}{a_0}\right)\equiv \y(\alpha), 
\ee
where $a_0$ is an integration constants, hence we set 
\be
a(t)=a_0 \,\alpha(t)
\ee
and we also set  the lapse function to the constant value, 
 \be
 N^2=\frac{12n^2}{m^2}\left(\frac{a_f}{a_0}\right)^{4s-2n}.
 \ee
Eq.\eqref{eqn} then assumes the form of the energy conservation, 
\be                \label{cons}
\frac{\dot{\alpha}^2}{\alpha^2}
+\U(\alpha)={\cal E},
 \ee
with the ``potential energy" 
 \be                    \label{V}
 \U(\alpha)=\alpha^{4s-2n}\left[
 (2n-s-1)c(\alpha)^{2n-s}-3(s+1)c(\alpha)^{-s}
 \right],~~~~~
 \ee
and  the ``total energy"
 \be
 {\cal E}=\frac13\,N^2 (m^2\lambda_g+\bm{\Lambda}).
 \ee
 The sign of the ``total energy" is determined by that of $m^2\lambda_g+\bm{\Lambda}$
 while its absolute value depends on $a_0$ and hence can be 
 arbitrary. 
Solving \eqref{eqn3} yields $\y(\alpha)$, injecting which to \eqref{V} determines $\U(\alpha)$, and then 
\eqref{cons} determines $\alpha(t)$. It is clear  that $\alpha$ should be confined to the 
region where $\U(\alpha)\leq {\cal E}$. 

The solution of the problem contains several subcases, depending 
on values of the parameters $n,s$.  
Let us first consider cases where the amplitude $\y$ is constant. 

\underline{\underline{I.  $n=2(s+1)$.}}  The solution of \eqref{eqn3} is an arbitrary constant
that can be assumed to be positive, $\y=\y_0>0$. The  potential becomes 
\be                                      \label{V0}
\U=\frac{3(s+1)c_0^{-s}(c_0^{4(s+1)}-1)}{\alpha^4},
\ee
which can be positive or negative, depending on values of $s$ and $c_0$. 
If $\U$ is negative, then it describes an effective radiation 
mimicked by massive gravitons. The scale factor then evolves as in the universe containing 
a radiation and an  effective cosmological term mimicked by ${\cal E}$. 
If $\U>0$ then the massive gravitons mimic   a ``phantom radiation".
One has in this case $\U(\alpha)\to +\infty$ as $\alpha\to 0$ hence there is an infinite potential barrier near singularity so that 
the solution is a {\it bounce}: the universe first shrinks up to a minimal non-zero size,
then hits the potential barrier and expands again. 
It is worth noting  that the very existence of bounces indicates  that the 
Null Energy Condition is violated. 

At the same time, the solutions can be free of ghosts and \blue{gradient instabilities}. Indeed, choosing 
$c_0$ to be close to unity, the no-ghost condition $\zeta(c_0)>0$ 
expressed by \eqref{zzz} will have approximately the same solutions as for $c_0=1$,
corresponding to the interior of the ellipse in Fig.\ref{Fig1}.
The two sound speeds
will then be close to unity.

\underline{\underline{II. $n\neq 2(s+1)$,  $\y=const$.}} The solution of \eqref{eqn3} 
\blue{in this case} is 
\be                          \label{cc}
\y^{2n}=\frac{3s}{2n-s}\equiv \y_\ast^{2n}, 
\ee
where one should assume that either  $2n>s>0$ or $2n<s<0$ for $\y_\ast$ to be positive. 
One has 
\be
\U=-\frac{6n \y_\ast^{-s}}{2n-s}\, \alpha^{4s-2n}<0.
\ee
If $n=2s$ then $\U=\blue{-4}$ and the universe expands with a constant Hubble \blue{expansion} rate. 
If $n>2s$ then for small $\alpha$ the universe is dominated by an effective \blue{``fluid"} 
mimicked by $\U$ while for large $\alpha$ the potential  approaches zero and the universe expands 
with a constant Hubble \blue{expansion} rate $H^2={\cal E}\blue{/N^2}$ (which should be positive).  
If $n<2s$ then the \blue{``fluid term"} $\U$  
grows without bound as $\alpha \to\infty$. 

The no-ghost condition \eqref{UV} reduces to 
\be
\zeta(c_\ast)=-\frac{3s}{2n}>0,
\ee
which is impossible to fulfill 
since $n$ and $s$ should be either both positive or both negative \blue{for $c_\ast$ to be positive}. 
Therefore,  type II solutions always have ghost. 

\blue{\underline{\underline{III. $n\neq 2(s+1)$,  $\y\neq const$.}}} 
Let us now study solutions for which the amplitude $\y$ is not constant. 
The general solution of \eqref{eqn3} for $n\neq 2(s+1)$  is 
\be                     \label{gen}
\frac{(2n-s)\,\y^{2n}-3s  }{\y^{s+1}}=\left(\frac{a}{a_0}  \right)^{2(n-2s-2) }.
\ee
Several subcases are to be considered. The left hand side of this expression 
is positive in one of the following  cases,
\bea                                           \label{cases}
&&{\rm (a)}~s=2n<0;~~~~
{\rm (b)}~s=0, n>0;~~~~
{\rm (c)}~n>s/2>0, c^{2n}>3s/(2n-s);~~  \nonumber \\
&&{\rm (d)}~n<s/2<0, c^{2n}<|3s/(2n-s)|;~~~~
{\rm (e)}~n>s/2,~s<0. 
\eea
Each of these cases further splits into subcases depending on the signs of $n$, 
of $s+1$,
of $2n-s-1$, and of $n-2s-2$. This renders the
classification of solutions a bit tedious, but still manageable. 

\blue{\underline{IIIa.}} Let us first consider case (a) in \eqref{cases}, where $s=2n<0$. 
Then, properly redefining the constant $a_0$ in \eqref{gen}, 
one obtains 
\be                   \label{c}
\y^{2n+1}=\left(\frac{a}{a_0} \right)^{2(3n+2)}\equiv \alpha ^{2(3n+2)},
%~~~\Rightarrow ~~~
%\y=\alpha^{2\frac{3n+2}{2n+1}}\in[0,\infty)
\ee
hence, \blue{as $\alpha$ increases from zero to infinity,} $c$ either increases from zero to infinity or decreases from 
infinity to zero.  The potential is 
\be              \label{VV} 
\U(\alpha)=-\alpha^{6n}-3(2n+1)\alpha^{-\frac{2n}{2n+1}}. 
\ee
Depending on value of $n$ there are several subcases. 

\blue{IIIa-1: $-1/2<n<0$,  $s=2n$.} $\y(\alpha)$ increases \blue{as $\alpha$ increases}. 
The potential $\U(\alpha)$ is always negative and \blue{its absolute value} 
becomes large for $\alpha\to 0$ and for $\alpha\to\infty$. 

\blue{IIIa-2: $-2/3<n<-1/2$,  $s=2n$.} $\y(\alpha)$ decreases \blue{as $\alpha$ increases}. 
The potential $\U(\alpha)$ is qualitatively similar to the one
shown in panel B in Fig.\ref{Fig2} below. 
It 
is large and positive at small $\alpha$  while 
for $\alpha\to\infty$  it approaches zero from below, hence there is a minimal value 
$\U_{\rm min}=\U(\alpha_{\rm min})<0$  \blue{with $\alpha_{\rm min}=1$}. 
If 
${\cal E}=\U_{\rm min}$ then the system always rests at the  minimum
and the geometry  is flat \blue{with $c(\alpha_{\rm min})=1$}. 
If $\U_{\rm min} <{\cal E}<0$ then $\alpha(t)$ 
oscillates   around the value $\alpha_{\rm min}$ while $c(t)$ oscillates 
around $c(\alpha_{\rm min})=1$. 
For ${\cal E}\geq 0$ the solution is a bounce. 

\blue{IIIa-3: $n<-2/3$,  $s=2n$.} $\y(\alpha)$ increases 
\blue{as $\alpha$ increases}. 
The potential $\U(\alpha)$ 
is qualitatively similar to the one
shown in panel A in Fig.\ref{Fig2} below. 
It is large and negative for small $\alpha$, then passes through a maximal value 
$\U_{\rm max}=\U(\alpha_{\rm max})>0$ \blue{with $\alpha_{\rm max}=1$}, 
then approaches 
zero from above as $\alpha\to\infty$. 
Depending on value of ${\cal E}$, 
the motions in this potential correspond either to cosmologies with an
initial  
singularity or to non-singular bounces. If 
$\U_{\rm max}={\cal E}$ then 
there is a solution for which $\alpha(t)$ grows
from the constant value $\alpha_{\rm max}$ in the past to infinity 
in the future, 
hence the universe interpolates between the flat space
and de Sitter space. 

There are also  two boundary cases,   $n=-1/2$ and $n=-2/3$. 
For  $n=-1/2$ the scale factor $a$ should be constant, as seen from Eq.\eqref{eqn3},
hence this is a particular case of the Minkowski solutions. 
The $n=-2/3$ case corresponds to the intersection  with family I since one has then 
$n=2(s+1)$. 

For all type \blue{IIIa} solutions the no-ghost condition {\eqref{zzz}}
reduces to 
\be
\zeta(c)=-3\,\frac{2n+1}{2n}>0\,,
\ee
which is independent of $c$. This condition is fulfilled for solutions 
of type \blue{IIIa-1}, hence they are ghost-free. However, one finds then  that the sound speeds are imaginary, 
hence there are \blue{gradient instabilities}. 
The no-ghost condition is violated for
solutions of types \blue{IIIa-2 and IIIa-3}. 

\blue{\underline{IIIb.}} In case (b) in \eqref{cases}, for $s=0$ and $n>0$, \blue{the no-ghost condition \eqref{zzz} becomes}
\be
\zeta(c)=\frac{1-2n}{2n}\,c^{2n} \blue{>0}. 
\ee 
This is \blue{satisfied} for $n<1/2$, however, there are \blue{gradient instabilities} 
in this case \blue{as the coefficient of $C_S^2$ in \eqref{ccc0} is positive}. 

\blue{\underline{IIIc.}} Let us now consider case (c) in \eqref{cases}, with $n>s/2>0$. 
 Then, redefining the integration constant $a_0$, Eq.\eqref{gen} can be rewritten as 
\be
\frac{\y^{2n}-\y_\ast^{2n}}{\y^{s+1}}=\left(\frac{a}{a_0}  \right)^{2(n-2s-2) }\equiv \alpha^{2(n-2s-2) },
\ee
where $\y_\ast$ is the same as in \eqref{cc}, hence 
one should have $\y\in[\y_\ast,\infty)$ for the left hand side 
to be positive. 
There are again several cases to study
and one   finds  the following possibilities. 

\blue{IIIc-1: $2n>s>0$,  $n>2(s+1)$.} $\y(\alpha)$ increases from $\y_\ast$ 
to infinity; the potential
$\U(\alpha)$ has  a maximum as shown in panel A in Fig.\ref{Fig2}. 

\blue{IIIc-2: $2n>s>0$,  $\max\{(s+1)/2,~2s\}<n<2(s+1)$.} $\y(\alpha)$ decreases
from infinity to $\y_\ast$; the potential
$\U(\alpha)$ has  a minimum as shown in panel B in Fig.\ref{Fig2}.

\begin{figure}[th]
\hbox to \linewidth{ \hss
	\resizebox{8cm}{6cm}{\includegraphics{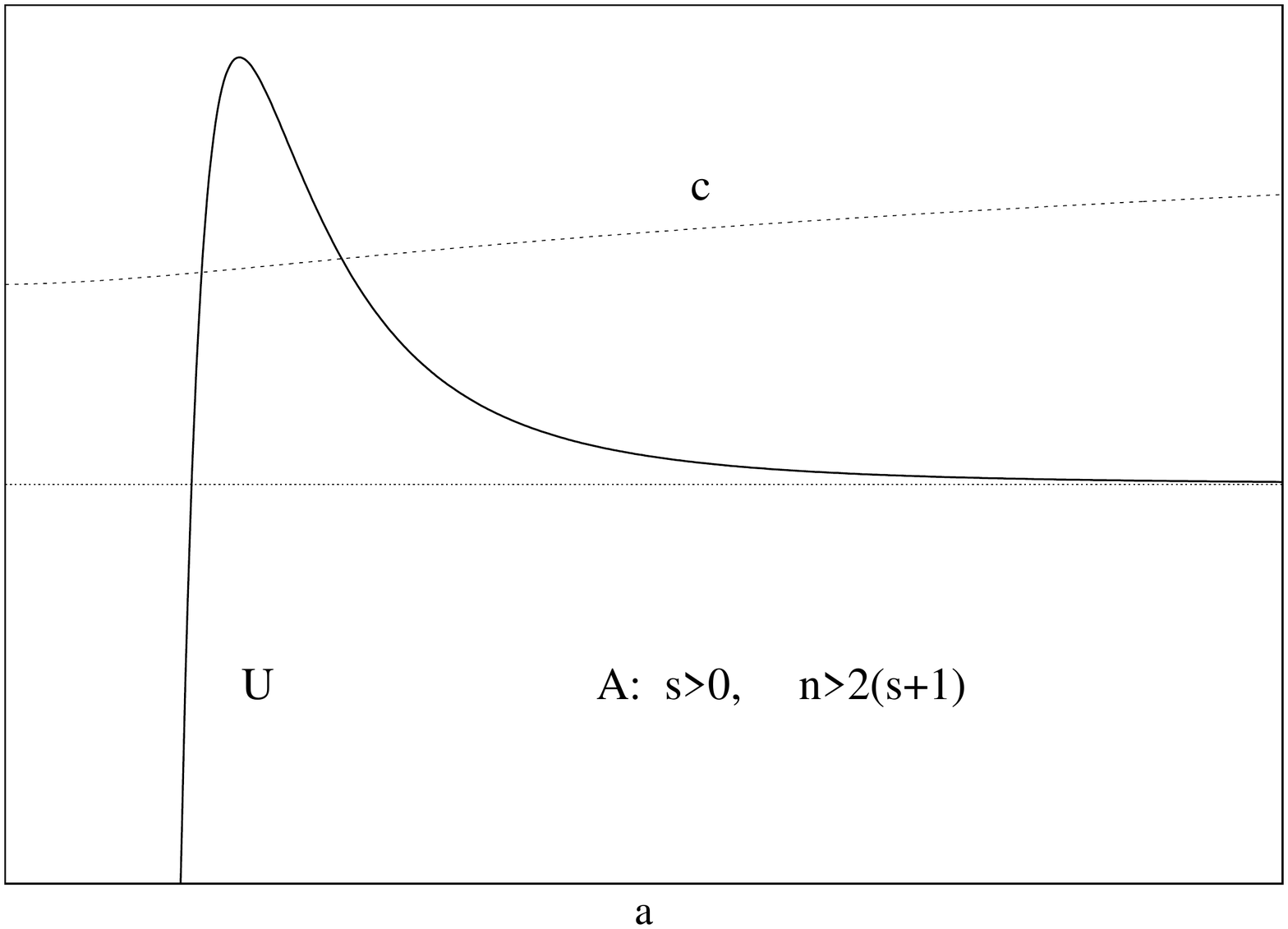}}
%\hspace{1mm}
	\resizebox{8cm}{6cm}{\includegraphics{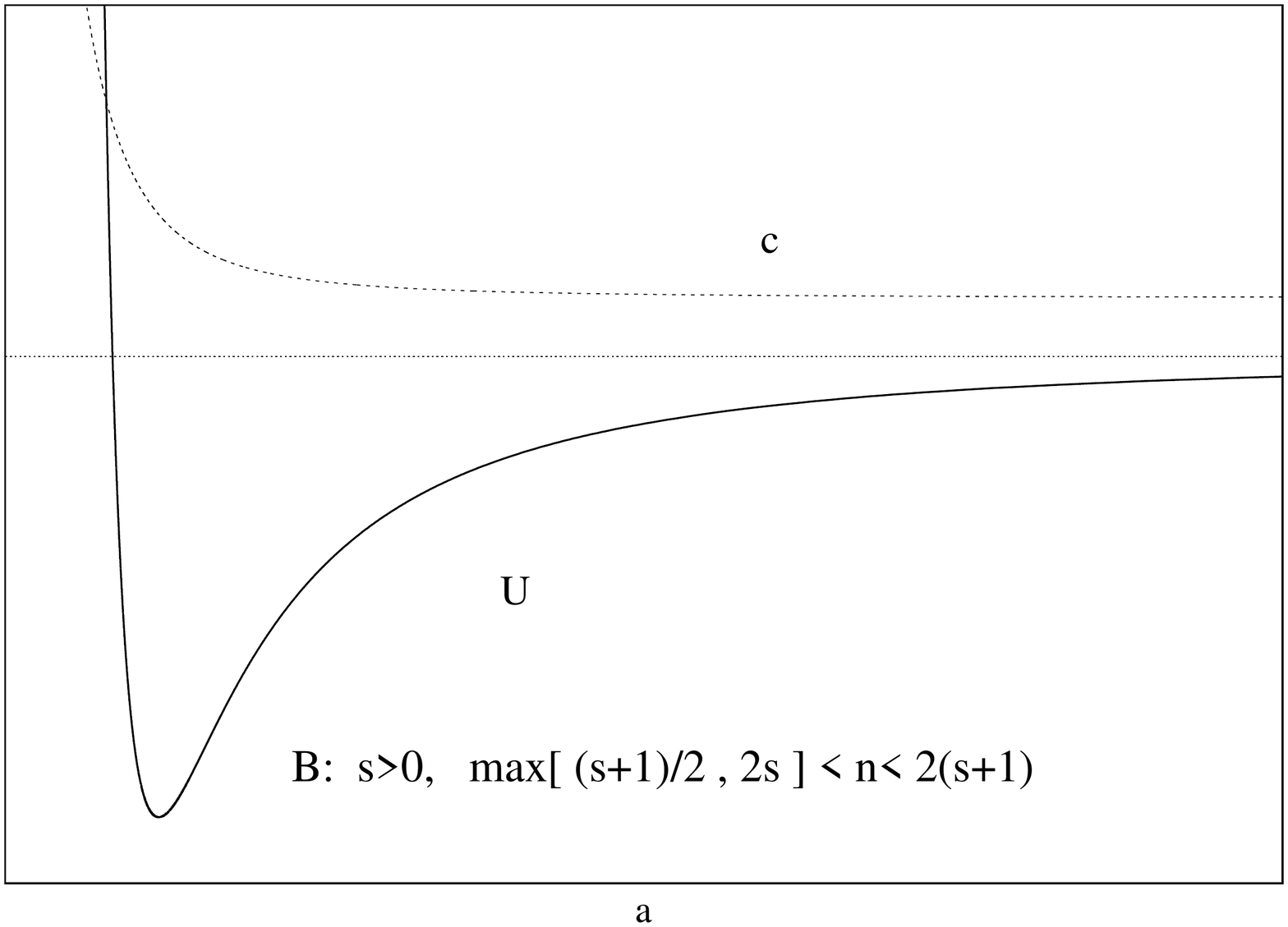}}
%\hspace{1mm}
\hss}
\hbox to \linewidth{ \hss
	\resizebox{8cm}{6cm}{\includegraphics{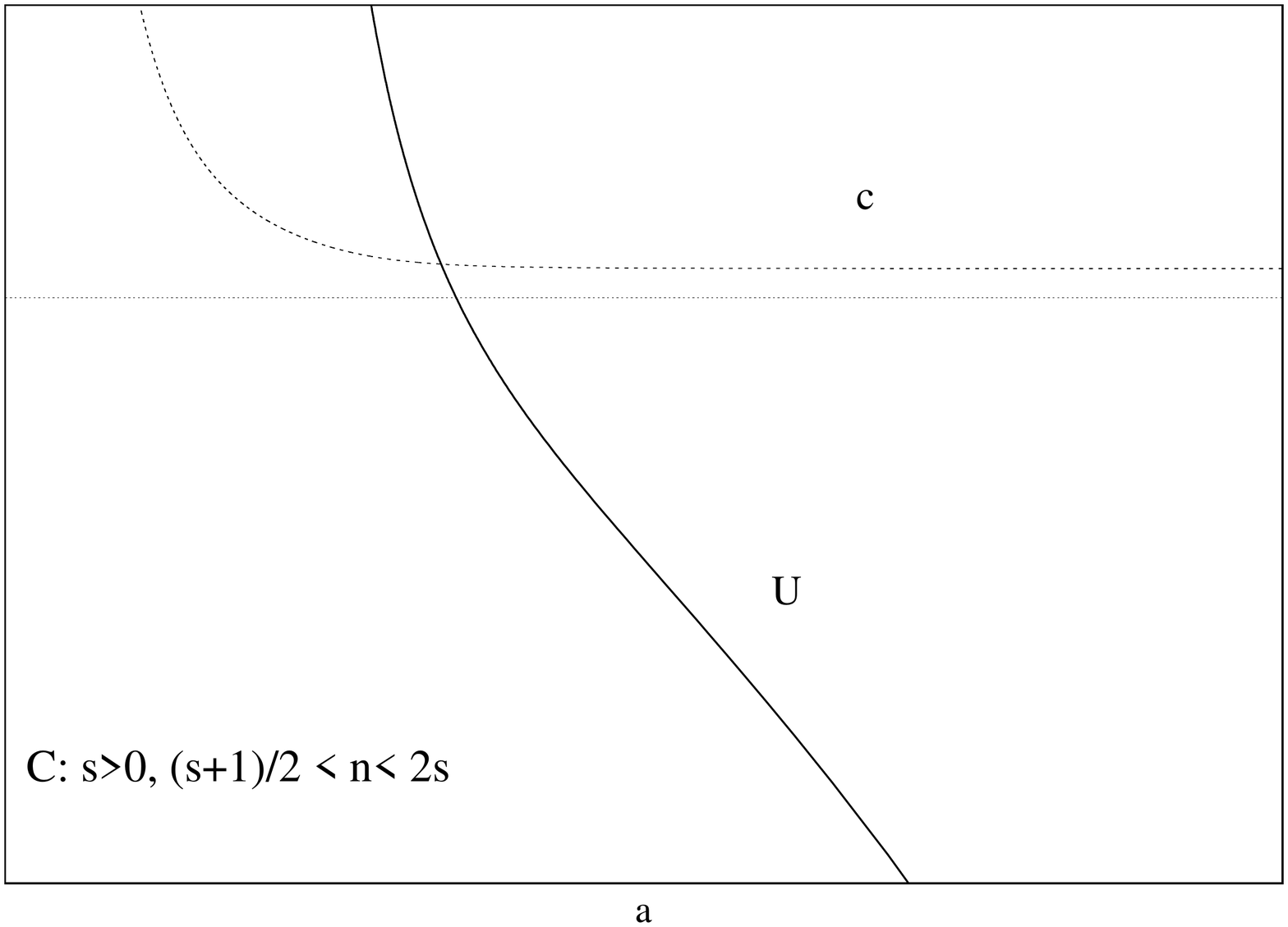}}
%\hspace{1mm}
	\resizebox{8cm}{6cm}{\includegraphics{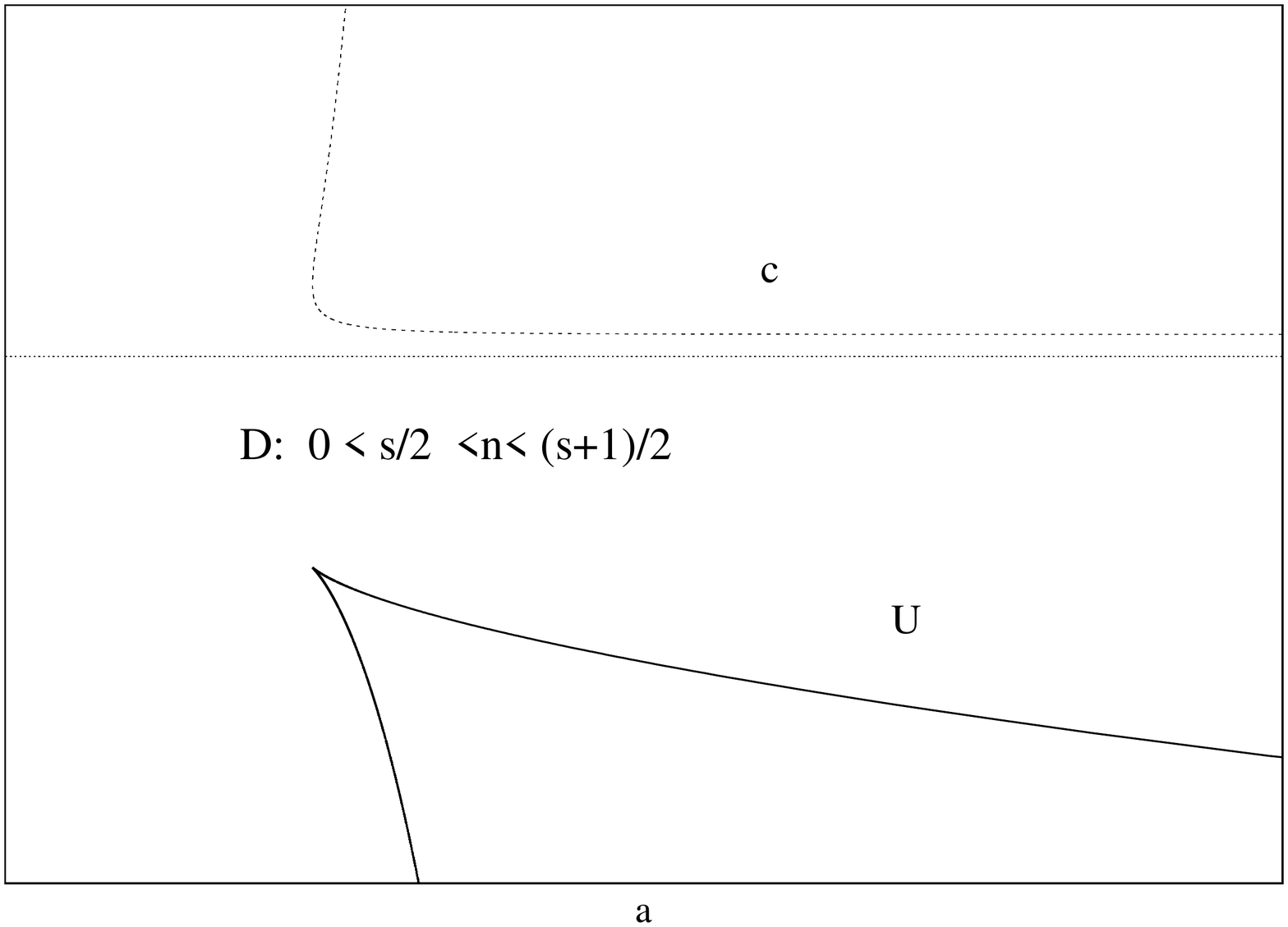}}
%\hspace{1mm}
\hss}
\caption{The potential $\U(\alpha)$ and $\y(\alpha)$ for generic $2n>s>0$.}
\label{Fig2}
\end{figure}

%\fi

\blue{IIIc-3: $2n>s>0$,  $(s+1)/2<n<2s$.} Both $\y(\alpha)$ and $\U(\alpha)$
decrease as shown in panel C in Fig.\ref{Fig2}. 

\blue{IIIc-4: $2n>s>0$,  $s/2<n<(s+1)/2$.} Both $\y(\alpha)$ and $\U(\alpha)$
are two-valued functions defined only for $\alpha\geq \alpha_{\rm min}>0$ 
as shown in panel D in Fig.\ref{Fig2}. For ${\cal E}<\U_{\rm max}$ there are two different bounce solutions 
corresponding to reflections from either the lower or upper branch of $U$. 

For solutions of types \blue{IIIc-1, IIIc-2, and IIIc-3}
the amplitude $c(\alpha)$ approaches the value 
$c_\ast$ either for small or for large $\alpha$, which insures that there is ghost 
\blue{since $\zeta(c_\ast)=-(3s)/(2n)<0$}
in these cases. 

The situation is more complex in case \blue{IIIc-4}. 
If one plots $a$, $\U$, and $\zeta(c)$ against $c$ in this case, one finds that 
$a(c)$ attains a minimal non-zero value at some $c_m$, 
 the potential $\U(c)$ attains
a maximum at the same time, while $\zeta(c)$ changes sign. As a result, both $c(\alpha)$ 
and $\U(\alpha)$ are double-valued functions as shown in panel D in Fig.\ref{Fig2}. 
Their two branches determine two different solutions
which should be considered independently. The upper branch of $\U(\alpha)$ corresponds 
to the lower branch of $c(\alpha)$ where one has $\zeta(c)<0$, 
hence there is ghost. The lower branch of   $\U(\alpha)$ corresponds to the upper 
branch of $c(\alpha)$ where $\zeta(c)>0$, hence this branch is ghost-free. 
However, one finds 
\blue{gradient instabilities} there.

\blue{\underline{IIId and IIIe.}} 
Nothing qualitatively  new is found  in cases (d) and (e) in \eqref{cases}. 
There are again several subcases to study, but each time one finds the potential to
be either of one of the types show in Fig.\ref{Fig2}, or of type \blue{IIIa-1}. 
All of these solutions contain ghosts and/or \blue{gradient instabilities}.

 This gives all solutions with the Minkowski reference metric.  
Only type I solutions can be stable. 
 
 \subsection{\blue{Solutions with Milne fiducial metric}}

\blue{If the fiducial metric is Milne, then $K=-1$, $N_f=\dot{a}_f$, and 
Eq.\eqref{eqn222a} reduces to }
\be
 \frac{\dot{\y}}{\y} +  [F_1(\y)-F_2(\y)]\,
 \frac{\dot{a}}{a}
  +F_2(\y)\,\y\,\frac{N}{a}  =0.                    \label{eqn222}
\ee
Setting $\y$ to a constant value yields 
\be                       \label{XXy}
\frac{\dot{a}}{N}=\frac{\y F_2(\y)}{F_2(\y)-F_1(\y)}, 
\ee
injecting which to \eqref{eqn} \blue{with $K=-1$} one obtains 
\be
3\left(\left[\frac{\y F_2(\y)}{F_2(\y)-F_1(\y)} \right]^2-1\right)\frac{1}{a^2}
=m^2\lambda_g+\bm{\Lambda}
-\frac{m^2}{4n^2}\, \y^{-s} \left[(2n-s-1)\, \y^{2n}-3\left(s+1\right)\right]\xi^{2n-4s}.
\ee
This equation determines $\xi=\xi(a)$, while \eqref{XXy},
imposing the gauge where $N=1$, insures that $a$ is a linear function of time. 
This specifies all functions in the problem. 
\blue{If $c\neq 1$ then 
the physical metric is not flat.}
If $c$ is close to unity then one can adjust $s,n$ such that the solution will be free 
of ghosts and tachyons (in the UV limit).

More general solutions with $\dot{c}\neq 0$ can be obtained by setting
\blue{$N=a$ in \eqref{eqn222}}, which gives the equation containing only $c$ and $a$. 
However, solutions of this equations are not immediately obvious. 
At the same time, as we learned above, solutions with a non-trivial $c$ usually 
show ghosts and/or \blue{gradient instabilities}. 

At this point,  we terminate our analysis of the OP theory and come to the conclusions. 

\section{Summary and concluding remarks}

We presented above our analysis of the theory constructed by 
Ogievetsky and Polubarinov in \cite{OP}. 
Taking apart the remarkable way it was obtained, 
it is just a bimetric massive gravity with a specially designed mass term. 
As any other bimetric theory, it 
has two tensors, $X^\mu_{~\nu}$ and  $Y^\mu_{~\nu}$, obtained 
by varying with respect to the two metrics, respectively. 
If the reference metric is flat, then one has on-shell $\partial_\mu Y^\mu_{~\nu}=0$. 
The specialty of the OP theory is that, by the non-linear field redefinition \eqref{g-g1}, its physical metric 
can be algebraically expressed in terms of the ``graviton field"   $\hh^{\mu\nu}$ 
in such a way that $Y^\mu_{~\nu}$ becomes {\it linear} in $\hh^{\mu\nu}$. 
Therefore, the field equations imply as a consequence 
the linear ``spin limitation condition" 
$\partial_\mu \Psi^\mu_{~\nu}+q\,\partial_\nu \Psi^\alpha_{~\alpha}=0$. 
In the OP's view, this property is very important as it allows one to carry 
out a ``clean" classification of the spin states present in the theory. 
The linear subsidiary condition arising on-shell is the 
key property that distinguishes the OP theory 
among  other bimetric models.

It is, however, not immediately  obvious what this property gives  in practical terms. 
The most striking point is that the OP theory has a non-zero intersection
with the  dRGT ghost-free massive gravity. However, for generic parameter 
values it propagates 6 degrees of freedom and shows ghost on flat background. 
According to the currently adopted view, these properties are unacceptable. 
Nevertheless, we studied the  phenomenology of the theory to see if 
something interesting may emerge. 

We found many different types of homogeneous and isotropic cosmological solutions,
including self-accelerating cosmologies, bounces, oscillating solutions, etc. 
Unfortunately, most of them show ghost and/or \blue{gradient instabilities}.
Surprizingly, however, we find  that the de Sitter space is stable
in a large region of the parameter space, in spite of the presence of the 
6-th degree of freedom, hence the Boulware-Deser mode becomes benign. 
Moreover, even the instability of the flat space does not seem 
so dramatic, as it turns out that the Milne universe, which is a sector of 
Minkowski space, is actually UV stable in the theory. Therefore, the instability
can only be due to the soft modes with wavelengths of the order or larger than the 
Milne horizon. This suggests that the flat space ghost instability is similar 
to the classical Jeans instability 
with respect to long-wave perturbations. Therefore,  the BD mode 
could probably 
be viewed as some kind of non-relativistic fluid \cite{Gumrukcuoglu:2016jbh}.

Our conclusion is that the OP theory does show interesting features, despite the presence of the 
6-th polarization. This  may be  due to the 
``spin limitation" encoded in the theory, even though there are other 
massive gravities with a healthy 6-th mode \cite{Celoria:2017bbh}. 
It would be interesting to study also other solutions in the theory, as for example black holes. 
It seems also that the particular OP theory with $n=1/2$ and $s=0$, which 
shows both the  ``spin limitation" and the ``freedom of the ghost", should be further
studied.

 \section*{Acknowledgements}

The work of S.M. was supported by Japan Society for the Promotion of Science (JSPS)  Grants-in-Aid for Scientific Research (KAKENHI) No. 17H02890, No. 17H06359, and by 
World Premier International Research Center Initiative (WPI), MEXT, Japan.  
M.S.V. thanks for hospitality the YITP in Kyoto, where a part of this work  was completed. 
Discussions during the workshop YITP-T-17-02  ``Gravity and Cosmology 2018" and the 
YKIS2018a symposium ``General Relativity -- The Next Generation" were useful.
His work  was also partly supported by the Russian Government Program of Competitive Growth 
of the Kazan Federal University.

\appendix
\setcounter{section}{0}
\setcounter{equation}{0}
\setcounter{subsection}{0}
\section{Stability of more general cosmologies \label{AppA}}

\renewcommand{\theequation}{\Alph{section}.\arabic{equation}} 
In this Appendix we derive the stability conditions for 
cosmological solutions described by metrics  \eqref{gf}. 
These metrics determine the matrices 
\be
S^\mu_{~\nu}=\xi^2\left(
\begin{matrix} 
y^2 & 0 \\
0 & \mathbbm{1}_3 \\
\end{matrix}
\right),~~~~~~
\Sigma^\mu_{~\nu}=(\hat{S}^n)^\mu_{~\nu}=
\xi^{2n}\left(
\begin{matrix} 
y^{2n} & 0 \\
0 & \mathbbm{1}_3 \\
\end{matrix}
\right).
\ee
Perturbing the metrics via 
\be                     \label{pert1}
 g_{\mu\nu}\to g_{\mu\nu}+h_{\mu\nu},~~~~~~~f_{\mu\nu}\to f_{\mu\nu}\,,
 \ee
 one has (with $i,j=1,2,3$)
 \be                         \label{dS}
 \delta S^\mu_{~\nu}=-g^{\mu\alpha}h_{\alpha\sigma}S^\sigma_{~\nu}\equiv 
 \left(
\begin{matrix} 
\delta S^0_{~0} & \delta S^0_{~j} \\
\delta S^i_{~0} & \delta S^i_{~j} \\
\end{matrix}
\right).
 \ee
 This does not commute with $S^\mu_{~\nu}$, which complicates the computation 
of $\delta\Sigma^\mu_{~\nu}=\delta (\hat{S}^n)^\mu_{~\nu}$. 
However, a direct calculation yields 
   \be                \label{pow}
 \hat{S}^k\,\delta \hat{S}\, \hat{S}^m=
\xi^{2k+2m}  \left(
\begin{matrix} 
y^{2m+2k}~\delta S^0_{~0} & ~~y^{2k}~ \delta S^0_{~j} \\
y^{2m} ~\delta S^i_{~0} & ~~\delta S^i_{~j} \\
\end{matrix}
\right).
 \ee 
 Assuming for a moment $n$ 
to be integer, one has 
\be
\delta\hat{\Sigma}=\delta(\hat{S}^n)=\delta\hat{S}\, \hat{S}^{n-1}+\hat{S}\,\delta\hat{S}\, \hat{S}^{n-2}+
\ldots +\hat{S}^{n-1}\,\delta\hat{S}, 
\ee
 and using \eqref{pow} one can sum up the geometric series, which yields 
 \be                \label{pow1}
 \delta\Sigma^\mu_{~\nu}=
\xi^{2n-2}  \left(
\begin{matrix} 
n\,y^{2n-2}~\delta S^0_{~0} & ~~\gamma~ \delta S^0_{~j} \\
\gamma ~\delta S^i_{~0} & ~~n\,~\delta S^i_{~j} \\
\end{matrix}
\right)~~~~~\mbox{with}~~~~\gamma=\frac{y^{2n}-1}{y^2-1}. 
 \ee 
 We now simply extend this expression  to arbitrary real values of $n$  
 and inject it to 
 \be                 \label{X1}
 \delta X^\mu_{~\nu}=\frac{s}{2}\,h\,X^\mu_{~\nu}+
 \frac{1}{2n^2}\,\phi^s\left(
 n\,\delta\Sigma^\mu_{~\nu}-\frac{s+1}{2}\,
 \delta \Sigma^\alpha_{~\alpha}\,\delta^\mu_{~\nu}
 \right)
 \ee
 to compute 
 \be                \label{X2}
 \delta X_{\mu\nu}=h_{\mu\alpha}\,X^\alpha_{~\nu}+g_{\mu\alpha}X^\alpha_{~\nu}\,,
 \ee
 which is to  be used in the perturbation equations \eqref{pert}. 
 %\be                \label{X3}
 %E_{\mu\nu}\equiv \delta G_{\mu\nu}+(\bm{\Lambda}+m^2\lambda_g)-m^2\delta X_{\mu\nu}=0. 
 %\ee
 We use the representation  \eqref{h} of $h_{\mu\nu}$ in terms of the scalar, vector, and tensor modes
 to insert  to \eqref{dS} to obtain $\delta S^\mu_{~\nu}$, which is then used in 
 \eqref{pow} and in \eqref{X1}--\eqref{X2} to obtain $E_{\mu\nu}$, and finally  in \eqref{act0}
 to obtain the effective action.
 
 The effective action splits again into the sum of the tensor, vector, and scalar parts. 
 In the tensor sector we obtain 
 \be                 \label{ITa}
 I_{\rm T}=\frac{M_{\rm Pl}^2}{4}\int \left(  
 \frac{1}{N^2}\left(\dot{D}_{1}^2+\dot{D}_{2}^2\right)-\left[M^2_{\rm T}+\frac{p^2}{a^2}  \right](D_{1}^2+D_{2}^2)
   \right)Na^3\,dt\, d^3x,
 \ee
 which is completely similar to the expression \eqref{IT} we had before, the only difference being the 
 change in the mass parameter value,
 \be
 M^2=m^2\xi^{2n-4s}\to  M^2_{\rm T}=m^2\xi^{2n-4s}\y^{-s},
 \ee
 which becomes a function of time since $\xi$ and $\y$ are now time-dependent. 
 
 In the vector sector we obtain 
  \be                     \label{IVa}
 I_{\rm V}=\frac{M_{\rm Pl}^2}{4\mu^2}\int  _{\rm T} \left(  
 \frac{p^2}{N^2[C_{\rm V}^2\, p^2/a^2+M^2_{\rm T}\,]}\left(\dot{V}_{1}^2+\dot{V}_{2}^2\right)-p^2(V_{1}^2+V_{2}^2)
   \right) Na^3\,dt\, d^3x
 \ee
 with 
 \be
 C_{\rm V}^2=n\,\frac{\y^2-1}{\y^{2n}-1}. 
 \ee
 This action  reduces to \eqref{IV} when $\y\to 1$. % since $\lim_{y\to 1}c^2(y)=1$.
 
 In the scalar sector we find
\be        
 I_{\rm S}=\frac{M_{\rm Pl}^2}{4}\int M^2_{\rm T}\sum_{a,b=1,2} \left(  
 \frac{1}{N^2}\,K_{ab}\,\dot{S}_a\dot{S}_b
 +\frac{\cal Q}{N}\,\epsilon_{ab}\,\dot{S}_a S_b - U_{ab}\,S_a S_b
   \right)Na^3\, dt\, d^3x, 
 \ee
 where the expressions for $K_{ab}$, ${\cal Q}$, and $U_{ab}$ are rather complicated,
 but  they simplify for $p\to\infty$. We obtain 
 \be                     \label{KKK}
 K_{11}=\frac{\zeta(\y)}{2{\cal H}^2}+{\cal O}(1/p^2),~~~~K_{22}=\frac{a^2p^2}{4\mu^4C^2_{\rm V}}+{\cal O}(1),
 ~~~~K_{12}=-\frac{\zeta(\y)}{C^2_{\rm V}}\frac{M_{\rm T}^2 a^2}{8\mu^2{\cal H}^2}+{\cal O}(1/p^2),
 \ee
 where ${\cal H}\equiv  {\dot{a}}/(Na)$ is the Hubble parameter of the background metric and 
\be                  \label{zzzz}
 \zeta(\y)\equiv -\frac{(2n-s-1)(2n-s)\y^{2n}+3s(s+1) }{4n^2}. 
 \ee 
%  The $K_{ab}$ components  in \eqref{KKK} reduce  to those in \eqref{KK} for $\y\to 1$. 
One has 
 \bea
 {\rm tr}(K_{ab})=\frac{a^2p^2}{4\mu^4C^2_{\rm V}}+{\cal O}(1), ~~~~~~
 \det(K_{ab})=\frac{\zeta(\y)\,a^2 p^2}{8\mu^4 C^2_{\rm V}{\cal H}^2}+{\cal O}(1),
 \eea
 and the positivity of these expressions, which guarantees the absence of ghost, yields 
  a single condition,
  \be                             \label{ZETA}
  \zeta(\y)>0.
  \ee  
This explains Eq.\eqref{UV} in the main text. 

One can also  compute the sound speed using \eqref{speed}. The necessary  elements are 
 \be
 {\cal Q}=-\frac{\omega}{4\mu^2\cal H}\, p^2+{\cal O}(1), 
 \ee
where
 \be
 \omega=-\frac{1+\y^2}{2C^2_{\rm V}}+\frac{2s+1}{2n}\,(1+\y^{2n})-\frac{s(s+1)}{2n^2}(3+\y^{2n})
 ~~~\stackrel{\y\to 1}{\longrightarrow}~~~~1+2\zeta. 
 \ee
 One also uses 
 \be
 U_{11}=-\frac{\y^2\,p^2}{4 a^2C^2_{\rm V}{\cal H}^2}+{\cal O}(1), ~~~~~
 U_{22}=-\frac{Z}{2\mu^4}\,p^4+{\cal O}(p^2),~~~~~~U_{12}={\cal O}(p^2),
 \ee
 where
 \be
 Z=-\frac{1}{4n^2}\left[2n(2n-2s-1)+s(s+1)(3+\y^{2n})  \right]~~~\stackrel{\y\to 1}{\longrightarrow}~~~~\zeta.
 \ee
 This is enough to compute the sound speed in the UV limit. 
 Eq.\eqref{speed} yields in the leading in $p$ order the equation for $C^2_{\rm S}$,
 \be                                               \label{ccc1}
 2C^2_{\rm V}\zeta(\y)\,C_{\rm S}^4+\left[C^4_{\rm V}\left(4Z\zeta(\y)-\omega^2 \right)+\y^2\right] C_{\rm S}^2
 +2\y^2C^2_{\rm V}Z=0,
 \ee
which gives rise to  Eq.\eqref{ccc0} in the main text. 

\blue{
We obtained the above conditions assuming that the spatial curvature in \eqref{gf} vanishes, $K=0$. 
However, since they are derived in the UV limit, they must be sensitive only to the local physics, which 
is the same for all values of  $K$, hence they are expected to be valid for any $K$. 
Rigorously speaking, it is possible  that the dimensionless combination $(m^2\lambda_g + \bm{\Lambda} - m^2 X^0_{~0})N^2a^2/\dot{a}^2-3$ ($=3N^2K/\dot{a}^2$), which vanishes for $K=0$, might appear in the no-ghost condition and the sound speed. Nonetheless, explicit calculations with $K\ne 0$ show that this is not the case. As a result, the no-ghost condition \eqref{zzz} 
and the algebraic equation \eqref{ccc0} that determines the sound speed are valid for any $K$.}

 %\section{Energy}
 %\setcounter{equation}{0}

%\bibliographystyle{JHEP}
%\bibliography{W}

\providecommand{\href}[2]{#2}\begingroup\raggedright\endgroup

\end{document}